%% file: main.tex
\documentclass[sigconf]{acmart}

\usepackage{listings}
\usepackage{xcolor}
\usepackage{colortbl}
\usepackage{algorithm}
\usepackage{algpseudocode}

\definecolor{codegreen}{rgb}{0,0.6,0}
\definecolor{codegray}{rgb}{0.5,0.5,0.5}
\definecolor{codepurple}{rgb}{0.58,0,0.82}
\definecolor{backcolour}{rgb}{0.95,0.95,0.92}

\AtBeginDocument{%
  \providecommand\BibTeX{{%
    \normalfont B\kern-0.5em{\scshape i\kern-0.25em b}\kern-0.8em\TeX}}}

\copyrightyear{2025}
\acmYear{2025}
\acmConference[SCF '25]{ACM Symposium on Computational Fabrication}{November 20--21, 2025}{Cambridge, MA, USA}
\acmBooktitle{ACM Symposium on Computational Fabrication (SCF '25), November 20--21, 2025, Cambridge, MA, USA}
\acmDOI{10.1145/3745778.3766655}
\acmISBN{979-8-4007-2034-5/2025/11}

\newif\ifsubmit
\submitfalse
\ifsubmit
    \newcommand{\stefanie}[1]{}
    \newcommand{\megan}[1]{}
    \newcommand{\mackenzie}[1]{}
    \newcommand{\faraz}[1]{}
    \newcommand{\liane}[1]{}
    \newcommand{\changes}[1]{}

\else
    \newcommand{\stefanie}[1]{{\leavevmode\color[rgb]{1.0, 0.0, 0.5}{#1}}}
    \newcommand{\megan}[1]{{\leavevmode\color[rgb]{1.0, 0.0, 0.2}{#1}}}
    \newcommand{\mackenzie}[1]{{\leavevmode\color[rgb]{1.0, 0.6, 0.0}{#1}}}
    \newcommand{\faraz}[1]{{\leavevmode\color[rgb]{0.0, 0.0, 0.0}{#1}}}
    \newcommand{\changes}[1]{{\leavevmode\color[rgb]{0.0, 0.0, 0.0}{#1}}}
    \newcommand{\liane}[1]{{\leavevmode\color[rgb]{0, 0.7, 1.0}{#1}}}
\fi

\usepackage{soul}

\usepackage{booktabs}
\usepackage{multirow}
\usepackage{graphicx}
\usepackage{microtype}

\begin{document}

\title[MechStyle]{MechStyle: Augmenting Generative AI with Mechanical Simulation to Create Stylized and Structurally Viable 3D Models}

\input{Sections/author_list}

\begin{abstract}

Recent developments in Generative AI enable creators to stylize 3D models based on text prompts. These methods change the 3D~model geometry, which can compromise the model's structural integrity once fabricated. We present MechStyle, a system that enables creators to stylize 3D printable models while preserving their structural integrity. MechStyle accomplishes this by augmenting the Generative AI-based stylization process with feedback from a Finite Element Analysis (FEA) simulation. As the stylization process modifies the geometry to approximate the desired style, feedback from the FEA simulation reduces modifications to regions with increased stress. We evaluate the effectiveness of FEA simulation feedback in the augmented stylization process by comparing three stylization control strategies. We also investigate the time efficiency of our approach by comparing three adaptive scheduling strategies. Finally, we demonstrate MechStyle's user interface that allows users to generate stylized and structurally viable 3D models and provide five example applications.

 \end{abstract}

\ccsdesc[500]{Human-centered computing~Human computer interaction (HCI)}


\keywords{personal fabrication; 3d printing; mechanical simulation; generative AI. }


\begin{teaserfigure}
\centering
  \includegraphics[width=\textwidth]{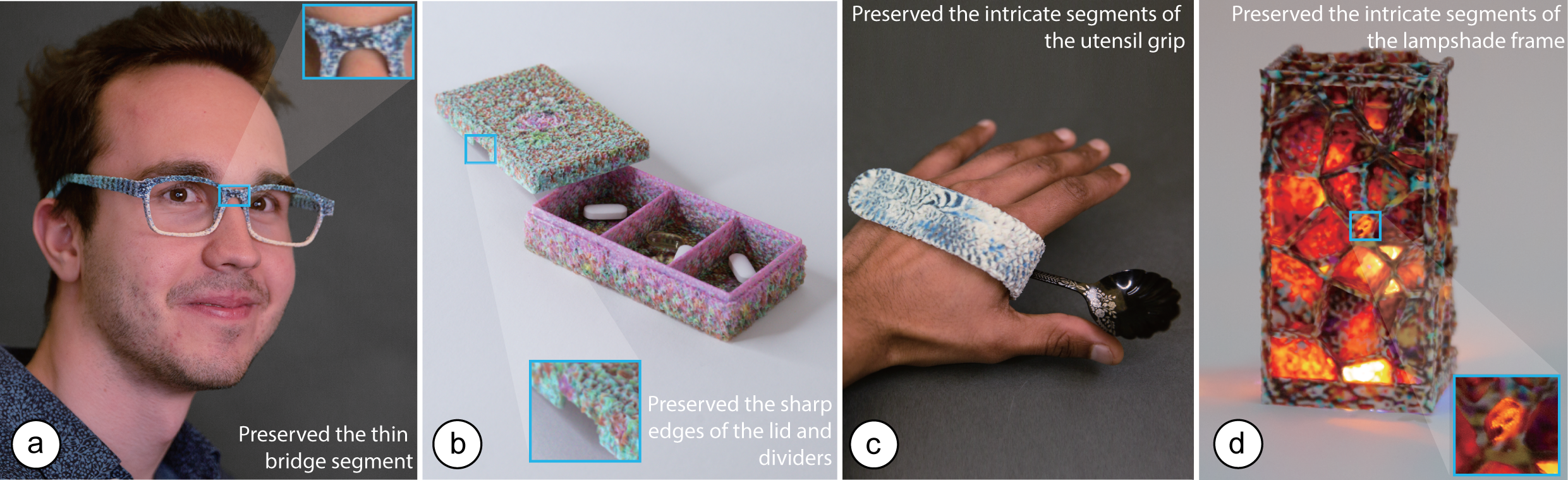}
  \caption{MechStyle enables creators to stylize 3D models with text prompts while preserving their structural integrity. Here, we used MechStyle to stylize five 3D models while ensuring that the printed objects do not break when accidentally dropped by the user: a)~an eye-glass frame stylized with `\textit{fish scales}' while preserving the fragile bridge. b)~a pill box with \textit{floral patterns}' that retains the sharp edges of the lid and dividers, c)~a utensil grip for users with fine motor impairments stylized as `\textit{Damascus steel}' pattern of the user's silverware set, d)~lampshade featuring `\textit{colorful mandala texture on mahogany wood}' while preserving its intricate structural segments. }
  \label{fig:teaser}
\end{teaserfigure}

\maketitle

\input{Sections/01_introduction}

\input{Sections/02_related_work}
\input{Sections/03_formative_study}

\input{Sections/04_mechanical_property_analysis}
\input{Sections/05_system}
\input{Sections/06_evaluation}

\input{Sections/07_implementation}
\input{Sections/08_applications}

\input{Sections/09_discussion}

\input{Sections/10_conclusion}

\input{Sections/questions}
\begin{acks}
\changes{We would like to extend our sincere gratitude to the MIT-Google Program for Computing Innovation for their generous support, which made this research possible. Furthermore, we thank Federico Tombari and Fabian Manhardt from Google Research for their valuable insights and feedback on this research.}
\end{acks}

\bibliographystyle{ACM-Reference-Format}
\bibliography{references}
\end{document}

%% file: Sections/author_list.tex
\author{Faraz Faruqi}
\email{ffaruqi@mit.edu}
\affiliation{%
  \institution{MIT CSAIL}
  \city{Cambridge}
  \state{MA}
  \country{USA}
}

\author{Amira Abdel-Rahman}
\email{amira.abdel-rahman@cba.mit.edu}
\affiliation{%
  \institution{Center for Bits and Atoms, MIT}
  \city{Cambridge}
  \state{MA}
  \country{USA}
}

\author{Leandra Tejedor}
\email{leandra0@mit.edu}
\affiliation{%
  \institution{MIT CSAIL}
  \city{Cambridge}
  \state{MA}
  \country{USA}
}

\author{Martin Nisser}
\email{nisser@uw.edu}
\affiliation{%
  \institution{University of Washington}
  \city{Seattle}
  \state{WA}
  \country{USA}
}

\author{Jiaji Li}
\email{jiaji@mit.edu}
\affiliation{%
  \institution{MIT CSAIL}
  \city{Cambridge}
  \state{MA}
  \country{USA}
}

\author{Vrushank Phadnis}
\email{vrushank@google.com}
\affiliation{%
  \institution{Google Research}
  \city{Mountain View}
  \state{CA}
  \country{USA}
}


\author{Varun Jampani}
\email{varun.jampani@stability.ai}
\affiliation{%
  \institution{Stability AI}
  \city{Boston}
  \state{MA}
  \country{USA}
}

\author{Neil Gershenfeld}
\email{neil.gershenfeld@cba.mit.edu}
\affiliation{%
  \institution{Center for Bits and Atoms, MIT}
  \city{Cambridge}
  \state{MA}
  \country{USA}
}

\author{Megan Hofmann}
\email{m.hofmann@northeastern.edu}
\affiliation{%
  \institution{Khoury College of Computer Sciences, Northeastern University}
  \city{Boston}
  \state{MA}
  \country{USA}
}

\author{Stefanie Mueller}
\email{stefmue@mit.edu}
\affiliation{%
  \institution{MIT CSAIL}
  \city{Cambridge}
  \state{MA}
  \country{USA}
}

\renewcommand{\shortauthors}{Faruqi et al.}

%% file: Sections/01_introduction.tex
\section{Introduction}
Generative AI is a rapidly growing field that democratizes creative expression. Co-creation tools based on Generative AI allow users to create new content by leveraging Deep Learning methods to learn from existing data in the form of text~\cite{chowdhery2023palm}, images~\cite{rombach2022high}, music~\cite{AI-music}, and 3D models~\cite{gao2022get3d}. Users prompt the system with a description of their desired output (e.g., “a vase with colorful flowers”), and the system returns a matching result. So far, these AI co-creation tools have mainly focused on creating digital content, such as images, text, and 3D models displayed on a screen. However, these results rarely hold up to the constraints of the physical world, limiting their applications in digital fabrication.

Unlike purely digital models, 3D models designed for fabrication are designed to accommodate both physical constraints and aesthetic goals---often with a complex interrelationship between the two. To reduce complexity, many novice creators use existing open-source designs as a starting point since those are assumed to be physically viable~\cite{Kuznetsov_2010_expertamateur,All-in-One-Print,X-Bridges}. Generative AI methods can support designers in modifying these 3D models based on text and image prompts. \faraz{In the context of 3D mesh editing, `style' refers to the visual and geometric attributes that define an object’s appearance. Prior work~\cite{michel2022text2mesh, x_mesh} in Generative AI methods have defined style modifications as encompassing color and local geometric modifications that enhance the object’s aesthetic appeal.} However, these methods do not consider the physical characteristics of the original design and as a result customized models may meet aesthetic goals but no longer have the physical viability of the original design. 

Researchers have proposed a variety of ways to highlight violations of physical properties for novice creators. For instance, Style2Fab~\cite{faruqi2023style2fab} maintains physical functionality when users stylize a 3D model using Generative AI by first separating functional and aesthetic parts in a model, and then only stylizing the aesthetic parts. This method maintains the model's physical affordances while approximating the desired style. However, physical properties, such as structural strength, are impacted by modifications on both the functional and aesthetic segments. Due to this complex relationship between form and function, it is essential to holistically analyze the impact a Generative AI method has on the physical properties of the model.   

Researchers have proposed tools that analyze 3D printable models to automatically identify weak segments~\cite{stochastic_structural_analysis} and subsequently fix them using hollowing, thickening, and strut insertion~\cite{stress_relief}. However, using such geometry changes to repair the model after it has already been stylized may distort the model's appearance and can conflict with the style intended by the user. On the other hand, 3D model stylization itself can be modified to optimize both the desired style and the desired structural properties. We argue that by converting stylization into a multi-objective optimization~\cite{liao2023interaction, multiple_objectives_uist_23}, we can preserve the structural integrity of the model while maximizing the user's desired style on the 3D model. By leveraging the iterative nature of the 3D manipulations during stylization, we can incorporate structural loss into the stylization process to optimize the geometry both for visual appeal and structural integrity.

In this paper, we present MechStyle, a method to stylize 3D~printable models while ensuring their structural viability. MechStyle accomplishes this by augmenting the iterative stylization process with structural feedback from a mechanical simulation. Thus, while the Generative AI model modifies the geometry to approximate the desired visual style, it uses structural analysis checks to reduce modifications on parts with increased structural stress. In this paper, we show how local structural stresses extracted from FEA (Finite Element Analysis) simulation can be used to inform the per-vertex manipulation. To efficiently run the augmented generative AI stylization process, we investigate how to effectively run mechanical simulation during the stylization process to ensure structural integrity while reducing run-time. We then integrate MechStyle as a user interface into an existing 3D editor to allow users to explore the trade-off between the desired aesthetic appearance (style loss) and the structural properties of the 3D model (structural loss).

\vspace{5pt}
\noindent In summary, we contribute an approach to augment generative AI stylization with mechanical simulation that ensures the structural viability of stylized fabricatable 3D models. This approach is built on the following system components:
\begin{itemize}
\item A method to incorporate structural stress feedback from a mechanical simulation into the Generative AI stylization process, leading to a combined loss function that optimizes style and structural strength;
\item Three stylization control strategies to reduce the impact of stylization in regions of the 3D model geometry with increased mechanical stress;
\item Three adaptive scheduling strategies for simulating intermediate stylized versions of the 3D model while reducing the overall run-time of the system.

\end{itemize}

%% file: Sections/02_related_work.tex
\section{Related Work}


To situate our findings and proposed system, we draw upon research on open-source 3D designs, methods to constrain Generative AI towards specific design goals, and stress analysis for 3D printing.

\subsection{Modifying Existing 3D Models}
As 3D printing has gained popularity, it has become the go-to fabrication tool of choice for a diverse group of makers. This trend has facilitated the growth of online repositories like Thingiverse\footnote{https://www.thingiverse.com/}, which serve as a source of re-usable models~\cite{alcock2016barriers, faruqi2021slicehub}. However, studies reveal that makers often encounter challenges when modifying existing 3D designs~\cite{hudson2016understanding, norouzi2021making, berman2021howdiy, schmidt2013design}. One primary obstacle is the complexity of current Computer-Aided Design (CAD) workflows, which are complex and thus require significant effort and time to learn. This challenge is amplified depending on the format of models shared online. Mesh files, such as those in STL/OBJ formats, are hard to edit in a semantically meaningful manner since they lack the meta-data associated with common CAD formats. Oehlberg et al.~\cite{oehlberg2015patterns} note that even when designs are labeled as customizable, they frequently fall short of the full spectrum of modifiability that makers desire.



To create custom functionality with existing 3D models, researchers proposed several strategies to work within the mesh editing constraints. For instance, PARTS \cite{hofmann2018greater} provides a method for designers to label their geometry with custom programs, making the models more reusable by novice makers. Alternatively, AutoConnect~\cite{koyama2015autoconnect} allows users to connect two meshes without modifying them by creating a 3D printable connector as a new mesh. Grafter~\cite{roumen2018grafter} creates new mechanisms from existing parts by scaling the meshes to fit the mechanisms without affecting their functionality. Attribit~\cite{chaudhuri2013attribit} allows makers to edit parts of their models with a pre-processed set of semantic attributes, such as making the head of a figure more ``scary'', which replaces the specific section with another previously labeled mesh with that attribute. In our work, we also investigate how to modify existing meshes but use Generative AI to perform local manipulations of the geometry to approximate a desired style based on text prompts. 

\subsection{Stylization with Generative AI}
Recent Generative AI tools, such as Text2Mesh~\cite{michel2022text2mesh} and TextDeformer~\cite{gao2023textdeformer} allow users to stylize existing meshes by changing their color and surface texture using text and image prompts. For instance, designers can stylize a vase 3D model to resemble a ``terracotta vase''. 3DHighlighter~\cite{decatur20233d} allows users to identify semantic regions on 3D models using textual descriptions, such as ``ears of a bunny'' or ``necklace on a human model'', which can be used to locally apply stylization on specific parts of the model. Applying textures during fabrication has also been explored directly at the G-code level. Yan et al.~\cite{yan2021man} introduce an on-the-fly method for embedding fine-scale geometric textures by modifying printer toolpaths, enabling fabrication of dense and natural-looking surface details. However, these methods do not consider functional or physical properties of a 3D~model. 

\subsection{Preserving or Generating Functionality in 3D Generative AI}

Researchers have proposed methods that augment Generative AI tools to create more compelling 3D models by incorporating different functionality goals when generating 3D~models~\cite{faruqi2024shaping}. For instance, while generating 3D~models from scratch, Mezghanni et al.~\cite{mezghanni2021physically, mezghanni2022physical} ensure that the model can stand upright by augmenting the generation process with a rigid-body simulation that checks for the stability of the model under gravity. Extending this to dynamic functionality, such as gripping and balancing, DiffuseBot~\cite{wang2024diffusebot} proposes a method for creating actuated soft-robot structures adept at these particular tasks. They choose their designs from a pre-trained sampling space of possible designs, simulate the design's performance in a differentiable simulator, and then use the feedback to choose an improved design from the sampling space. 

The above methods generate 3D models with the desired functionality from scratch. However, stylization is performed on existing 3D models, which necessitates the identification of existing physical functionality present in the model. Style2Fab~\cite{faruqi2023style2fab} is a stylization method that preserves functional regions in existing 3D models. It separates functional regions from aesthetic regions by training a classifier on functional segments derived from a manually labeled dataset of printable models. TactStyle~\cite{faruqi2025tactstyle} extends this line of work by enabling stylization of 3D models that incorporates not only visual appearance but also tactile properties. However, a model's structural properties, are impacted by modifications on both functional and aesthetic regions. Due to this complex relationship between form and function, it is essential to holistically analyze the impact a generative method has on the physical properties of the model. One standard technique to analyze physical properties is Finite Element Analysis (FEA), however integrating FEA into Generative AI methods poses multiple challenges.

\subsection{Stress Analysis for Creating Structurally-Viable 3D Models}

The standard computational technique for structural analysis of solid structures is FEA, specifically worst-case structural analysis~\cite{zhou_worst_case_structural}. This approximate method finds regions of high local stress or local deformations. Under specific conditions, structural analysis can be conducted without an FEA simulation. For instance, FastForce~\cite{abdullah2021fastforce} detects structural flaws without running an FEA simulation by identifying weakly connected regions in 3D models designed to be laser-cut. However, FastForce assumes that the structural flaws are due to poorly connected substructures in the geometry that can be identified using a graph, and only supports enclosed structures that are laser-cut. Additive manufacturing methods, such as 3D printing, can fabricate a wider range of geometries, with more complicated substructures. Thus, as a generalizable approach, researchers use FEA simulation to identify stress in 3D~models prior to fabricating them. Stava et al.~\cite{stress_relief} use the local stress information to create a structurally viable version by local thickening, hollowing, and strut insertion. Similarly, Zehnder et al.~\cite{zehnder2016designing} present a computational design tool for ornamental curve networks that combines user-driven aesthetics with structural eigenanalysis to ensure stability of delicate, filigree-like surface designs.  FabForms~\cite{shugrina2015fab} is a system that allows creators to explore structurally viable designs by pre-computing valid regions in a parametric design space using FEA and displaying the results in a user interface. Skouras et al.~\cite{skouras2013computational} present a method for creating actuated and deformable versions of 3D~models. They identify optimal locations for actuators in the model, and then optimize the model's geometry using FEA simulation to ensure they can reach each target pose.

However, given that FEA is computationally expensive, incorporating it into Generative AI methods at every iteration is computationally not feasible. In MechStyle we propose using simulations at targeted iterations of the stylization process that are most informative about changes in a model's structural properties.

%% file: Sections/03_formative_study.tex
\section{Formative Study}

\begin{figure*}[ht]
    \centering
    \includegraphics[width=\linewidth]{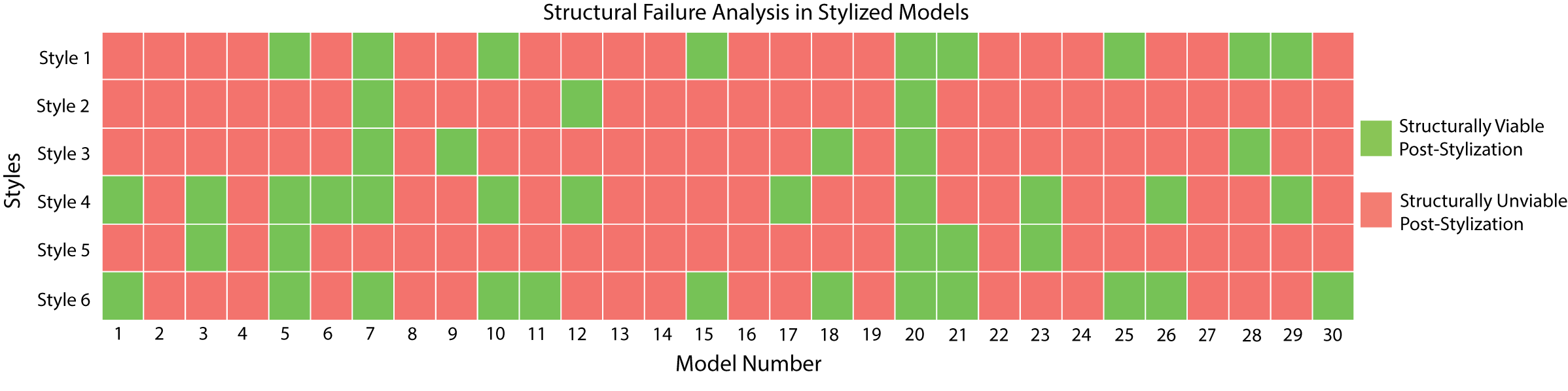}
    \vspace{-7mm}
    \caption{Formative Study Results: We stylized 30 popular 3D models downloaded from Thingiverse. Each model was stylized with 6~popular styles extracted from a stylization benchmark~\cite{x_mesh}. The stylized models were then simulated and analyzed for structural failure to identify models that are structurally unviable after fabrication. The model-style combinations colored in red were found to be structurally unviable post-stylization, while the model-style combinations colored in green were viable. Only 25.55\% of models were structurally viable after being stylized.}
    
    \label{fig:formative_study}
\end{figure*}

 We performed a formative study to investigate how Generative AI-based stylization affects the structural integrity of 3D models. Each combination of Generative AI stylization prompt and 3D model results in unique geometric changes, which are tightly associated with mechanical properties. Thus, in this formative study, we investigate the impact of stylization on a model's structural integrity by comparing the impact different styles have on the same 3D~model and the impact the same style has on different 3D~models. 

\paragraph{Dataset and Stylization:}
To investigate the impact of stylization across different 3D models, we generated 180 stylized 3D models (6~styles x 30~models). We chose the 6~most common stylization prompts from the X-Mesh dataset~\cite{x_mesh}, a benchmark for text-driven 3D stylization. The styles were (1) \textit{`brick'} , (2) \textit{`stone'}, (3) \textit{`cactus'}, (4) \textit{`realistic`}, (5) \textit{`crochet'}, and (6) \textit{`green'}. The X-Mesh dataset also contains 30~3D models but we did not use those because they consist of mostly decorative figurines (e.g., a dragon, a tiger) and are not representative of the mechanical complexity of models intended for fabrication. Instead, we collected the 30 most popular models from Thingiverse~\cite{Thingiverse2023Apr} that were structurally viable in their provided scale. \faraz{We standardized the models to have a resolution of 25K vertices, as outlined in Style2Fab~\cite{faruqi2023style2fab} for mesh standardization. The resolution of the model remained consistent during the stylization process. Next, both the original stylized models were converted into tetrahedral (TET) models for FEA simulation with fTetWild~\cite{hu2020fast}. The average number of elements across all TET models was 28.2k (std dev: 12k). The high standard deviation can be explained by the significant increase in complexity from the original model to the stylized model due to the geometrical changes during stylization. }


\paragraph{Finite Element Analysis Simulation Method:}
\faraz{To evaluate structural integrity, we perform a mechanical simulation of dropping the model onto a flat solid surface from a fixed height, following Langlois et al.'s~\cite{stochastic_structural_analysis} method. Mechanical simulations have been shown as effective methods in prior work~\cite{stress_relief, stochastic_structural_analysis} to identify weak regions in a 3D model. Specifically, Finite-Element based simulations are preferred over other heuristic approaches, such as geometric wall-thickness assessment, because a 3D model's structural viability is a product of several factors. These include not only the model's geometry but also the forces applied and the material used. Thus, a local heuristic-based approach might be ineffective, since it does not take into account all the forces applied to, and the mechanical properties of, a 3D model. Moreover, since stylization makes geometrical changes in a 3D model, they also influence the model's weight distribution, which has cascading effects on how forces are applied at local regions, affecting its structural viability. Thus, to capture all structural influences of stylization on a 3D model, we use mechanical simulation as our evaluation method.  We use the simulation environment WARP~\cite{warp2022}, and PLA as the material, which is a popular choice for 3D printing. }



\paragraph{Failure Analysis:}
After the simulated drop-test, we analyze if the model is still structurally viable using the failure analysis method from Stava et al.~\cite{stress_relief}. We use the von Mises yield criterion~\cite{mises1913mechanik}, which evaluates if the maximum absolute stress in the model after stylization is above the material's yield stress value (45.6 MPa for PLA). If the stress value is above the yield threshold, the model is no longer structurally viable and will likely break when dropped. 



\subsection{Results}
\autoref{fig:formative_study} shows the results of the failure analysis for all 30 3D models stylized with all 6 styles. The models highlighted in red are not structurally viable after stylization, while those highlighted in green retain their viability. Only 25.55\% of models are structurally viable after being stylized. This shows the significant impact Generative AI stylization methods have on the structural viability of models.

\begin{figure*}[h]
    \centering
    \includegraphics[width=0.6\linewidth]{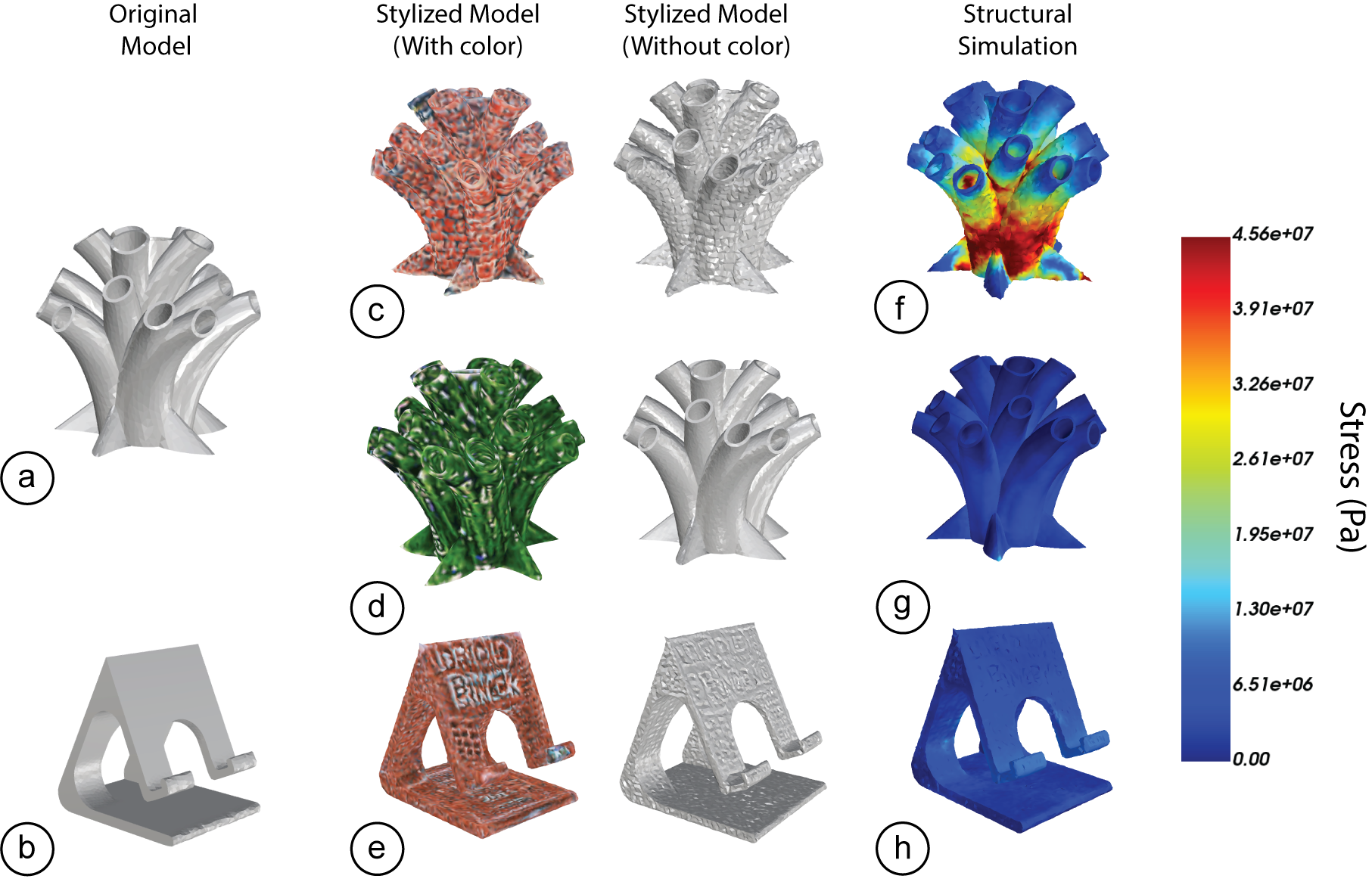}
    \vspace{-3mm}
    \caption{\faraz{Formative Study Results: Comparison of impacts of different styles on different 3D models. We find that different styles have different impacts on the same 3D model, and that different models are differently impacted by the same style. a)~The original pen holder 3D model. b)~The original phone stand 3D model. c)~Pen Holder stylized as a \textit{`brick pen holder'}. d)~Pen Holder stylized as \textit{`green'}. e)~Phone Stand stylized as a \textit{`brick phone stand'}. f)~Stress distribution for the \textit{`brick pen holder}' model. g)~Stress distribution for the \textit{`green}' pen holder. h)~Stress distribution for \textit{`brick phone stand}' model.}  }
    \label{fig:combined_models_styles}
\end{figure*}

\paragraph{Different styles have different impacts on the same 3D model.} 
\autoref{fig:combined_models_styles} shows the same pen-holder 3D model (Thing ID:73489) stylized as ``\textit{brick pen holder}'' (\autoref{fig:combined_models_styles}c), and as ``\textit{green pen holder}'' (\autoref{fig:combined_models_styles}d). The ``\textit{brick}'' (model number:~11, style number:~1 in ~\autoref{fig:formative_study}) style impacted the geometry such that the model was no longer structurally viable, with the maximum stress being 4~times the yield value of PLA. \autoref{fig:combined_models_styles}f shows the red region of the model where the simulated stress value equals or surpasses the maximum stress value of the material---this region would likely break when the object is dropped. \faraz{We can observe in the modified geometry view of the `pen-holder' (\autoref{fig:combined_models_styles}c), shown without color to highlight geometrical changes, that stylization introduced new ridge-like modifications on the model. Such changes may change its weight distribution, and reduce the strength of the lower segments of the model, which result in a stylized model with a significantly weakened base. For the same pen holder stylized as ``\textit{green}'' (\autoref{fig:combined_models_styles}d), the maximum stress was under the yield stress of PLA as shown by the blue highlighting in \autoref{fig:combined_models_styles}g (model number:~11, style number:~6 in ~\autoref{fig:formative_study}). This variation in maximum stresses experienced by differently stylized versions of the same model is due to how stylization prompts uniquely manipulate the geometry. In contrast, the `\textit{brick} style model, the same pen-holder when stylized as `green' does not undergo major geometrical changes, resulting in only small changes to the model's original structural strength. Thus, we find that different styles have different impacts on the same 3D model. }


\paragraph{Different models are differently impacted by the same style.}
We also found that the same style results in different maximum stress values for different 3D models. Recall that the `\textit{brick}' style rendered the pen-holder unviable (\autoref{fig:combined_models_styles}c,f). \faraz{However, when the same `\textit{brick}' style is applied to a model of a phone stand (Thing ID:2120591) as shown in \autoref{fig:combined_models_styles}e, it remains structurally viable (model number:~25 style number:~1 in ~\autoref{fig:formative_study}). We can observe that stylization introduced similar geometrical and color changes on both the models, but the structural simulation result of the two models shows significantly different stress distributions. Thus, different model geometries are differently impacted by stylization. In this example, the thin walls and complex overhangs of the pen holder are weakened more than the thicker segments of the phone holder.}


\vspace{5pt}
In summary, our formative study showed that stylizing 3D models significantly impacted their structural viability; only about 1 in 4 models was structurally viable post-stylization~(25.55\%). Additionally, we observed that the same model is impacted differently depending on the chosen style, and applying the same style across different models yields varying results. In the next sections, we present MechStyle, a system that utilizes in-situ simulation during the stylization process to preserve the model's structural integrity while approximating the desired style. 







%% file: Sections/04_mechanical_property_analysis.tex
\section{System Overview}
MechStyle integrates mechanical simulation into the 3D model stylization to systematically control the stylization process. As the model is stylized, MechStyle simulates intermediate versions of the model to identify local regions where the stress has increased, and then reduces stylization in subsequent iterations for those regions, preserving the functional geometry.

\subsection{MechStyle User Interface and Workflow}
\faraz{Users can access MechStyle through a user-interface plugin for the open-source 3D design software tool Blender~\cite{blenderUI}. To stylize a model with MechStyle, the user: (1)~loads the model, (2)~specifies their print material, and (3)~specifies the desired style as a text prompt. MechStyle then processes the model and standardizes it to have a 25k vertex resolution based on Style2Fab~\cite{faruqi2023style2fab}'s evaluation for stylization. Next, MechStyle stylizes the model while using the integrated simulation to preserve its structural integrity during the process. After stylization, the user can toggle between the stylized model and its FEA results, which are shown as a heatmap with red regions of increased stress. Finally, the user exports the stylized model for fabrication.

Since the wait-time of a MechStyle result can be significant, this UI tool also allows the user to rapidly explore different styles. Such a workflow would entail the following steps. Although the stylization choice is by default `MechStyle', the user can switch to `Freestyle' mode. In this mode, structural simulation feedback is not incorporated during the stylization process. The model is fully stylized based on the text prompt. Finally, the stylized model is simulated using MechStyle's FEA simulation tool to extract a heatmap. Using this mode allows the users to rapidly explore different styles and analyze its impact on the structural integrity of the model. Finally, after deciding on their intended style, the user can switch to MechStyle mode. Here, stylization is done with feedback from the structural simulation and the stylized result contains both the desired style of the user while maintaining its structural viability.} 

\begin{figure}[ht!]
    \centering
    \includegraphics[width=\linewidth]{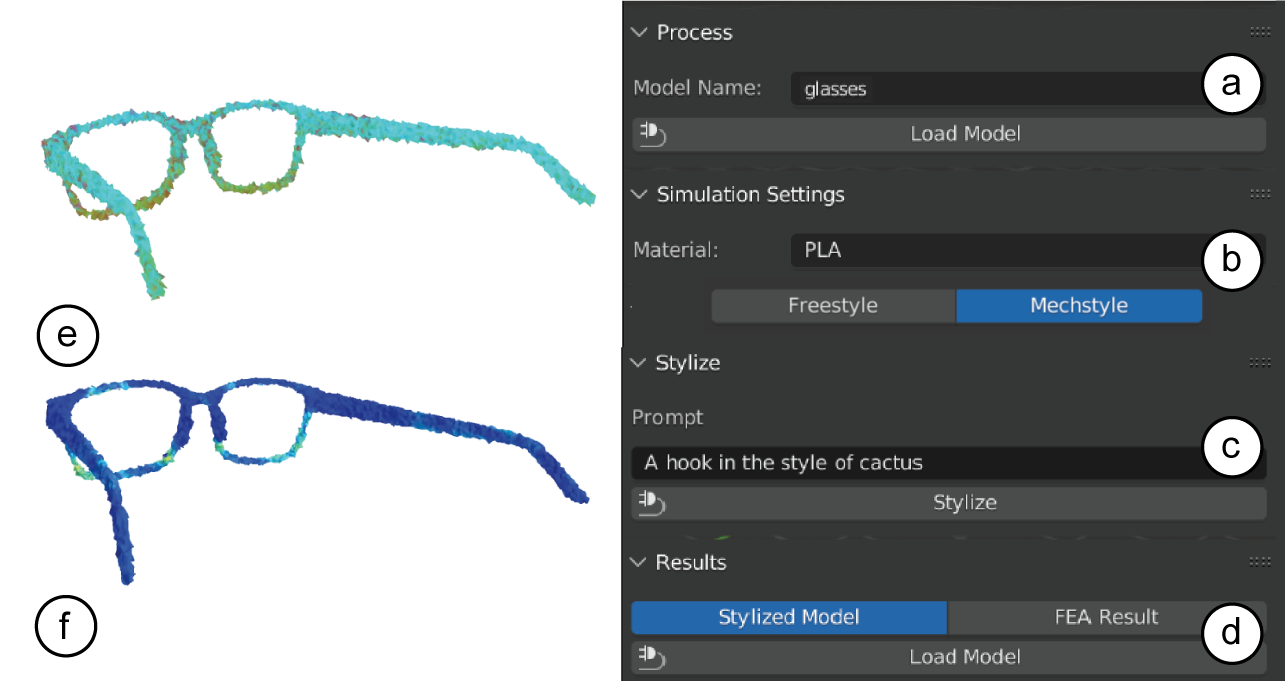}
    \caption{\faraz{In MechStyle's UI, the user a) loads a 3D model, b) enters their material for fabrication and chooses MechStyle or Freestyle technique, and c) enter their stylization prompt. d) the model is processed with the style, and the stylized model is loaded in the UI. e) Next, the UI allows the users to explore the stylized model, and also f) the FEA result of the stylized model to show regions of high stress, and its structural viability. }}
    \label{fig:mechstyle-ui}
\end{figure}

\subsection{Method}
We investigate two specific challenges for integrating mechanical simulation into the stylization process.
The first challenge is incorporating feedback from the mechanical simulation into the subsequent stylization steps to control the impact stylization has on regions of high stress. We present three \textit{stylization control strategies} for incorporating the stress results from a simulation of an intermediate stylized version of the model into the subsequent stylization process. 
The second challenge is reducing the number of computationally intense simulations that are run throughout the stylization process by identifying instances when simulations will be most informative. We present three \textit{adaptive scheduling strategies} to determine when the simulation should be executed to best preserve structural integrity.

\begin{figure*}[t]
    \centering
    \includegraphics[width=0.6\linewidth]{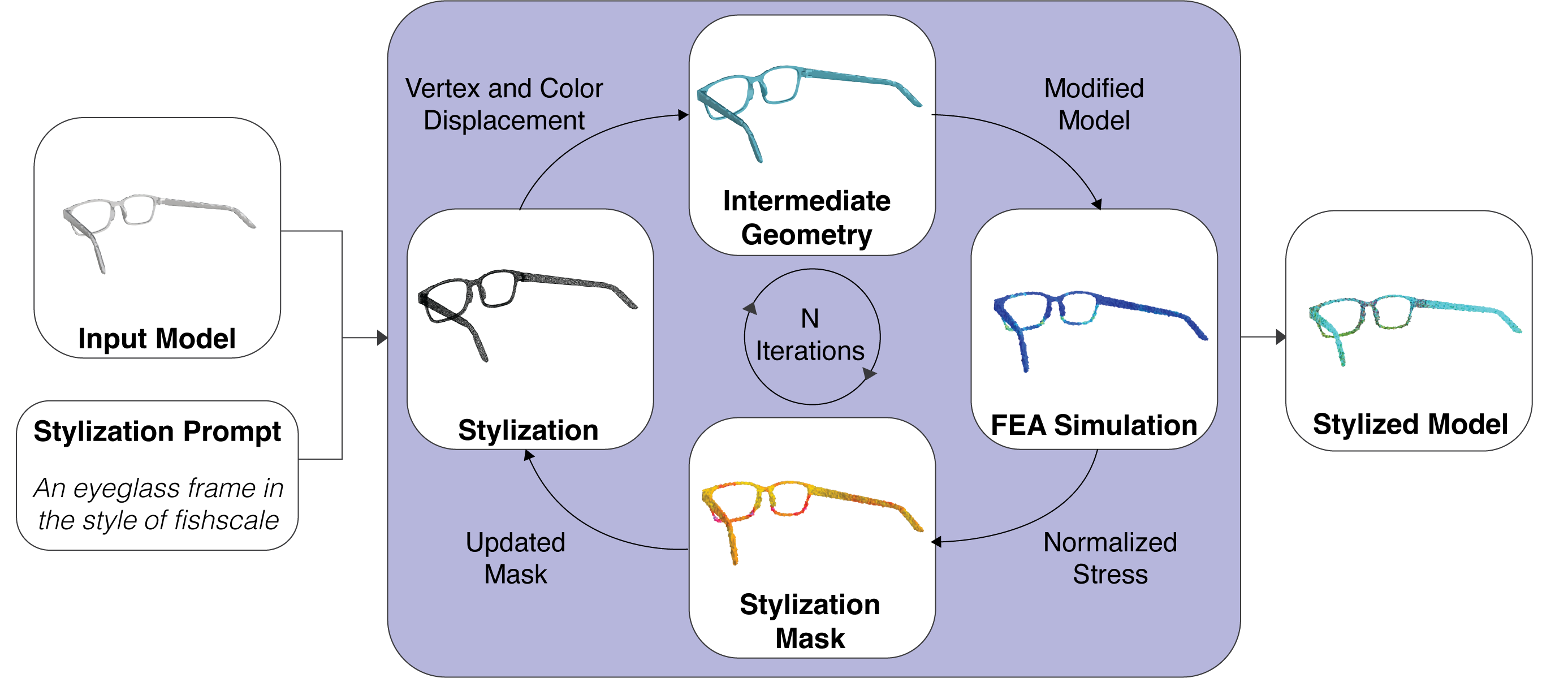}
    \caption{\faraz{MechStyle System Architecture. The stylization module iteratively modifies a model's geometry, while the simulation module simulates the intermediate stylized models to identify stress regions and update the stylization mask.} }
    \label{fig:system_figure}
\end{figure*}

To integrate stylization and simulation into one system, MechStyle is built on two components: the \textit{stylization module} and the \textit{simulation module}~(\autoref{fig:system_figure}). The stylization module takes the user's desired style prompt and iteratively modifies the geometry. Each iteration of stylization produces a new, unique intermediate 3D model. The simulation module tracks changes in stress on the model resulting from these intermediate models. Stylization occurs in every iteration, while simulation is only run on specific intermediate models. This reduces the time expended on simulation. As the stylization module modifies the geometry, MechStyle tracks the cumulative geometry and stress changes across iteration steps.  Thus, these two modules work in tandem, iteratively stylizing the model while controlling for the impact on structural integrity.


\subsection{Stylization Method} 
Stylization progresses by iteratively modifying a model with small displacements of its vertices and changes in their color. Over many iterations, these changes converge on a new model that matches the given style prompt. These changes are done gradually, with each vertex changing the model by <1\% of the model size after each iteration of stylization~\cite{x_mesh}.

We define a 3D model as an input mesh  $M$ with vertices in Cartesian coordinate $\mathbb{V}_M \in \mathbb{R}^{nx3}$ connected by faces $\mathbb{F}_M \in \{1,...n\}^{mx3}$ with color channels $\mathbb{C}_M \in \mathbb{R}^{nx3}$ corresponding to R, G, B colors. We use the X-Mesh~\cite{x_mesh} model architecture for stylization, which takes a text prompt from the user and iteratively displaces each model vertex and modifies their color channels to conform to the text prompt. Vertices are shifted along their normal to the mesh surface. The set $\mathbb{D} = \{d_v \forall v \in \mathbb{V}\}$ defines the distance along each normal the vertex will move in a given iteration. That is, stylization shifts each vertex $v$ by a distance of $d_v$ along the normal of $v$ relative to the mesh. This forms a modified 3D~model geometry. The color of the model is shifted with similar displacements in the color channel $\mathbb{C}$. We run this process for N~=~200 iterations, which has been shown to converge on a solution across a variety of models and styles~\cite{x_mesh}. This results in a stylized model $M_s$.

\subsection{Mechanical Simulation Method}\label{simulation-method}

We use mechanical simulation to estimate the stress on regions of the model caused by forces similar to those a model may be subjected to after fabrication. 
Our choice of simulation was informed by two constraints: the mechanical simulation should be able to support complex geometries and be computationally efficient to simulate several intermediate iterations of the stylization. Thus, we use a solid FEA-based method based on tetrahedral elements using a hyperelastic neo-hookean constitutive model~\cite{smith2018stable}. We implemented this using NVIDIA Warp~\cite{warp2022}, a Python framework for developing differentiable GPU-accelerated high-performance simulations. This simulation is material-specific and is initialized using a given material's Lamé parameters. The Lamé parameters, denoted as $\lambda$ and $\mu$, are material-specific constants that can be calculated using a material's Young's Modulus (E) and Poisson's ratio ($\nu$). For our paper, we focus on the widely used material PLA for 3D printing. Additional materials can be added by changing the Lamé parameters provided to the simulation. 

For the simulation, we use a model- and use-case agnostic stochastic FEA simulation technique previously proposed by Langlois et al.~\cite{stochastic_structural_analysis} that includes simulating a model under a Drop Test scenario. They simulate the model dropping on a hard surface from different heights and based on the forces the model is subjected to compute the stress distribution across the model. In our simulation, we set the environment to have an impenetrable surface, and then simulate the 3D~model dropping under gravity from a height of 1.5 meters, which is the same height that Langlois et al. ~\cite{stochastic_structural_analysis} use in their analysis. 


To conduct the failure analysis, we use the von-Mises yield criterion~\cite{mises1913mechanik} based on Stava et al.~\cite{stress_relief}. This method compares the von Mises stress values $\sigma_v$ for all $v$ vertices in the model $M$ with a material-specific yield strength $S_{material}$ for a material. We define the critical level of stress at which a model is likely to break as $\sigma_c$ (\autoref{eq:critical_value}). This critical value is defined by the Lamé parameter $\lambda$, which is a constant $0 < \lambda < 1$ that specifies the maximum allowed value of stress with respect to the yield strength of the material $S_{material}$. Decreasing the value of $\lambda$ increases the sensitivity of the method but can also lead to the generation of false positives. For our experiments, we use the value $\lambda = 0.2$, as specified in Stava et al.~\cite{stress_relief}. 

\begin{equation}\label{eq:critical_value}
    \sigma_c = \lambda \cdot S_{material}
\end{equation}

\section{Simulation-Informed Stylization via Masking} 
Each iteration of stylization produces an intermediate version of the stylized model. We identify local regions with increased stress by simulating this model version and conducting a failure analysis using Stava et al's~\cite{stress_relief} method. We propagate this information to the stylization module as a mask to reduce per-vertex stylization in areas of increased stress during subsequent stylization iterations. 

\paragraph{Identifying Regions of Increased Stress:}
For a specific iteration~$i$ of stylization, the original model $M_0$ is modified to create an intermediate geometry $M_{i}$ which is put through the mechanical simulation to extract the stress distribution $\mathbb{S}_{i}$. We do a failure analysis by normalizing $\mathbb{S}_i$ with $\sigma_c$, the material-specific critical stress value. This gives us a normalized stress vector $\mathbb{S}_{norm}(i)$, which is a measure of how close the vertices are to the critical stress value and, by extension, how close the material is to failure. For a vertex $v_i$ on the model $M_i$, the normalized stress value is $s_{norm}(v_i)$ (\autoref{eq:s_norm}). When $s_{norm}(v_i) > 1$, the stress on the model $M_i$ at that vertex has gone beyond the critical stress value and is no longer structurally viable. Conversely, when $s_{norm}(v_i) < 1$, the stress is under the critical stress value for the material and remains viable. 


\begin{equation}\label{eq:s_norm}
    s_{norm}(v_i) = \frac{s_{v_i}}{\sigma_c}
\end{equation}
\begin{equation}\label{eq:S_norm}
    \mathbb{S}_{norm}(i) = \{s_{norm}(v_i) \forall v_i \in \mathbb{V}_{M_i}\}
\end{equation}

\paragraph{Creating the Stylization Mask:}
Our goal is to keep the stress on each vertex $v$ of the model below the critical value~(i.e.,~$s_{norm}(v_i)~<~1$). We accomplish this by locally masking individual vertices $V$ to reduce their displacement during subsequent iterations of stylization. The masking function uses the normalized stress vector $\mathbb{S}_{norm}$ to reduce or freeze the geometry displacement on each individual vertex. A given masking function $mask(v)$ returns a weight between 0 and 1, which is applied to reduce the displacement distance of a vertex during stylization (i.e., $d_v * mask(v)$). When the mask is 1, the vertex can move freely for stylization. As the mask approaches zero, displacement is increasingly constrained at that vertex.
This produces the masked displacement set $\mathbb{D}_{mask}$ (\autoref{eq:d_norm}). We do not normalize the color modification values since they do not impact structural properties. 



\begin{equation} \label{eq:d_norm}
    \mathbb{D}_{mask} = \{d_{v_i} * mask(v_i) \forall v_i \in \mathbb{V}_{M_i}\}
\end{equation}

In the next section, we describe three stylization control strategies that apply the mask created by $\mathbb{S}_{norm}$ in different ways to reduce stylization in regions with increased stresses while staying close to the stylization prompt.

\subsection{Types of Stylization Control Strategies}\label{sec:masking-strategies}
We developed three different stylization control strategies that reduce stylization by different amounts in regions with increased local stresses. The first strategy linearly reduces per-vertex displacement according to the increase in stress. The second approach exponentially reduces the per-vertex stylization. The third approach freezes the stylization completely in those regions. 


\paragraph{Linearly Weighted Stylization:} In this approach, the displacement applied to each vertex during the stylization process is reduced linearly based on how close the local stress at that vertex is to the critical stress value. As the stress increases, the amount of stylization decreases proportionally.

 
\begin{equation}\label{eq:linear}
    mask_{v_i} = 1 - s_{norm}(v_i)
\end{equation}
%

\paragraph{Exponentially Weighted Stylization:} In this approach, we exponentially reduce the stylization in areas with increased stress. This technique is more sensitive to high stresses because as the model's stress values approach the yield stress value of the material, stylization gets more significantly reduced. To compute the mask, we calculate the negative exponent of $ s_{norm}(v_i) / 1 - (s_{norm}(v_i))$, which exponentially decreases as the $\mathbb{S}_{norm}$ approaches 0. This means the subsequent geometrical displacement is exponentially smaller as the stress increases. \autoref{eq:exponential} shows the exponential masking function for a given vertex. 

\begin{equation} \label{eq:exponential}
    mask_{v_i} =  e^{-\left(\frac{s_{norm}(v_i)}{1 - {s_{norm}(v_i)}}\right)}
\end{equation}

\begin{figure}[h]
    \centering
    \includegraphics[width=0.8\linewidth]{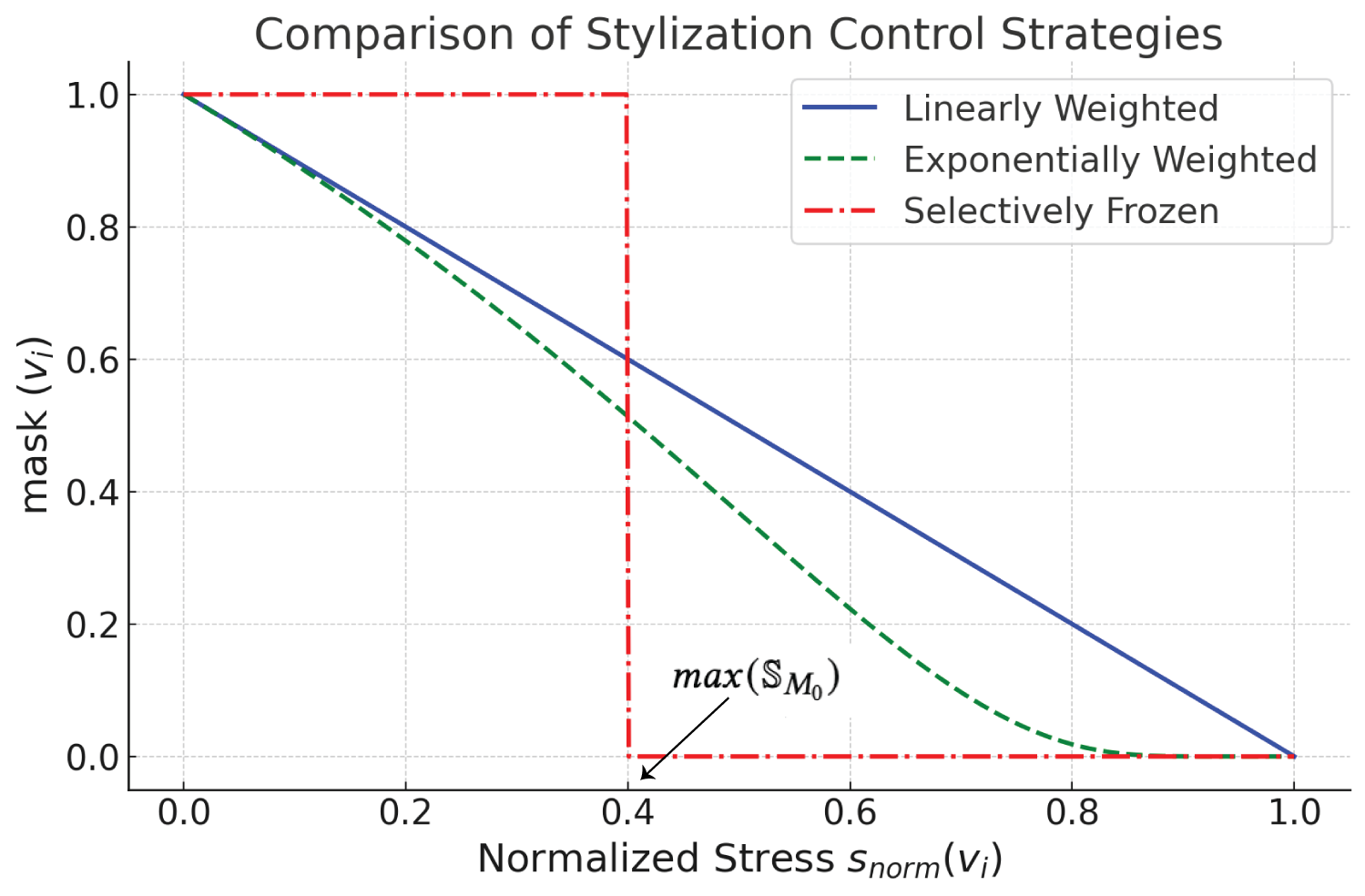}
    \caption{Visualization of three stylization control strategies. $\mathbb{S}_{norm}(i)$ is a measure of how close the individual vertices in the model are to the critical stress value $\sigma_{c}$. The ``Linearly Weighted'' strategy reduces the stylization effect proportionally as the local stress approaches $\sigma_{c}$. The ``Exponentially Weighted'' strategy reduces the stylization effect more sharply using an exponential function. The ``Selectively Frozen'' strategy completely stops the stylization once the stress reaches the maximum stress in the original model (max($\mathbb{S}_{M_0}$)). The y-axis, labeled as "mask \( (v_i) \)," represents the modulation applied to the displacement at each vertex \( d_i \).
}
    \vspace{-3mm}
    \label{fig:masking_strategies}
\end{figure}

\paragraph{Selectively Frozen Stylization:}
In this approach, the displacement matrix is immediately set to 0 once the vertices exceed a certain stress value resulting in no more vertex displacement in that region. Unlike the other approaches, selective freezing cannot use failure analysis (i.e., $\mathbb{S}_{norm}$) since once the failure criterion has been reached, subsequent iterations are likely to jump significantly past the failure point. Instead, while setting the stress threshold, we chose the maximum stress in the original model since the original model was structurally viable. This method is similar to that employed in Style2Fab~\cite{faruqi2023style2fab} where stylization is frozen in functional segments of the model.

\begin{equation}\label{eq:frozen}
    mask_{v_i}=\begin{cases}
            0& \text{if } s_{v_i} \geq max(\mathbb{S}_{M_0})\\
             1& \text{if } s_{v_i} <  max(\mathbb{S}_{M_0})
        \end{cases}
\end{equation}

In summary, we control the per-vertex stylization in areas with increased stresses by creating a mask based on the simulation results. We normalize the displacement vector $\mathbb{D}$ before applying it to the model $M_i$ with the updated mask. We compare these different stylization control strategies in our technical evaluation, which evaluates how closely the desired style was approximated (i.e., style-loss) and if the model was structurally viable after stylization.


\section{Adaptive FEA-Simulation Scheduling}\label{sec:scheduling-strategies}

Another challenge when trying to preserve the structural integrity of the model during stylization is to identify when feedback from the mechanical simulation will be most informative to the stylization process. A naive method would run a simulation at each iteration step of stylization. However, another important factor to consider for this integration is the total run-time of the combined system. FEA simulations are computationally intensive, and even though we leverage a GPU-accelerated simulation, running simulations at each iteration step is computationally prohibitive. We analyze the simulation run-time for stylized models and then discuss different scheduling strategies to efficiently run the simulation during stylization.  

\paragraph{Differences in Stylization and Simulation Times:}\label{simulation-stylization-times}
Ideally, FEA simulations would assess stress changes after each iteration of stylization. However, the run-time for a single FEA simulation is significantly more than that of one stylization iteration. Across the 30 models from our formative study dataset, one iteration of stylization with any style takes on average 2.67~seconds (std = 1.17), while the FEA simulation of the same models took an average of 4.61~minutes (std = 2.62). Since stylization runs for 200 iterations before convergence, simulating after each iteration increases the average run-time to 15.52 hours. Therefore, running FEA simulations at the same granularity as stylization is not feasible. 


Thus, the challenge here was to find an efficient scheduling strategy that reduces overall runtime while controlling for stylization's impact on the 3D model's structural viability. In the next section, we describe a set of scheduling techniques for simulation that can be incorporated into the stylization process.

\subsection{Types of Adaptive Scheduling Strategies}
\label{sec:scheduling_strategies}
To develop a method to determine at which stylization iterations simulation is the most informative, we developed three different adaptive scheduling strategies. The first strategy is time-based, i.e., it distributes simulation after a specific number of stylization iterations. The second strategy is geometry-based, i.e., it correlates the local iterative geometry change in the model with the local 3D~model thickness. The third strategy is stress-based, i.e., it correlates the local geometry change from stylization to the local stress, thereby monitoring geometry changes in structurally weaker areas more closely.


\paragraph{Temporal Scheduling:} Temporal scheduling runs simulation after a certain number of stylization iterations. We differentiate between linear and non-linear scheduling. For linear scheduling, simulations are uniformly distributed throughout the stylization process. However, a non-linear scheduling strategy might be more effective since stylization makes larger geometry changes at early iterations and then converges on more subtle changes \cite{x_mesh}. Moreover, displacements across iterations accumulate over iterations throughout the stylization. Thus, simulating more often early in the process captures stress changes related to those early geometry changes. Some examples of a non-uniform scheduling strategy are quadratic or exponential scheduling.


\paragraph{Geometry-Based Scheduling:}
This scheduling strategy determines when to run the FEA simulation based on the cumulative displacement of vertices over a number of stylization iterations with respect to the local model thickness. This approach is motivated by the tendency of models with thin geometries to be impacted more by stylization than thicker geometries. Thus, even small vertex displacements on thin geometries influence the structural integrity of the model. We track cumulative displacement at all vertices across iterations, and simulations are triggered when displacement surpasses a threshold based on the model thickness at that vertex. This heuristic method ensures that simulations are conducted in direct response to the stylization's impact on the geometry. In our experiments, we use a threshold of 10\% of local thickness. 



\paragraph{Stress-Based Scheduling:}
This strategy refines the geometry-based approach by accounting for local stress values. The local stress values in the original model provide information on which sections are structurally weak. We use the $\mathbb{S}_{norm}$ values to evaluate the model's initial structural viability. Combining this information with the geometric-thickness information (Geometry-Based Approach)  helps us determine when a vertex $v_i$ with high stress ($s_{norm}(v_i)$ close to 1) is undergoing a geometric change that could lead to an increase in stress. In this strategy, we assign vertex-level thresholds by modifying the geometry-based threshold proportionally with the local normalized stress value. By monitoring geometry changes in structurally weaker areas more closely, this method ensures that simulations are prioritized based on the potential risk to the model’s structural viability.



\vspace{5pt}
In summary, we designed three scheduling strategies that adaptively determine when to simulate an intermediate step of the stylization process to minimize run-time while preserving structural integrity. We evaluate these three different scheduling strategies in our technical evaluation for their run-time efficiency, ability to preserve structural integrity, and their ability to approximate the desired style.  

%% file: Sections/06_evaluation.tex
\section{Technical Evaluation}
We evaluated MechStyle to understand if it preserves the structural integrity of different 3D models given a variety of styles. We tested which of our stylization control strategies and adaptive scheduling strategies are most effective in preserving structural integrity, approximating desired style, and minimizing runtime. We used the same 30 models and 6 styles as in the formative study, resulting in 180 stylized models per condition. 

For stylization control strategies, we evaluate all three techniques described in section~\ref{sec:masking-strategies} (i.e., linearly-weighted, exponentially-weighted, selectively frozen). For adaptive scheduling techniques, we evaluate all four techniques described in section~\ref{sec:scheduling_strategies} (i.e., linear and exponential temporal scheduling, geometry-based, and stress-based), where the linear temporal scheduling strategy is the control condition. In total, we ran MechStyle in 2,160 different configurations (30 models x 6 styles x 3 stylization strategies x 4 scheduling strategies). 

\begin{figure*}[h]
    \centering
    \includegraphics[width=\linewidth]{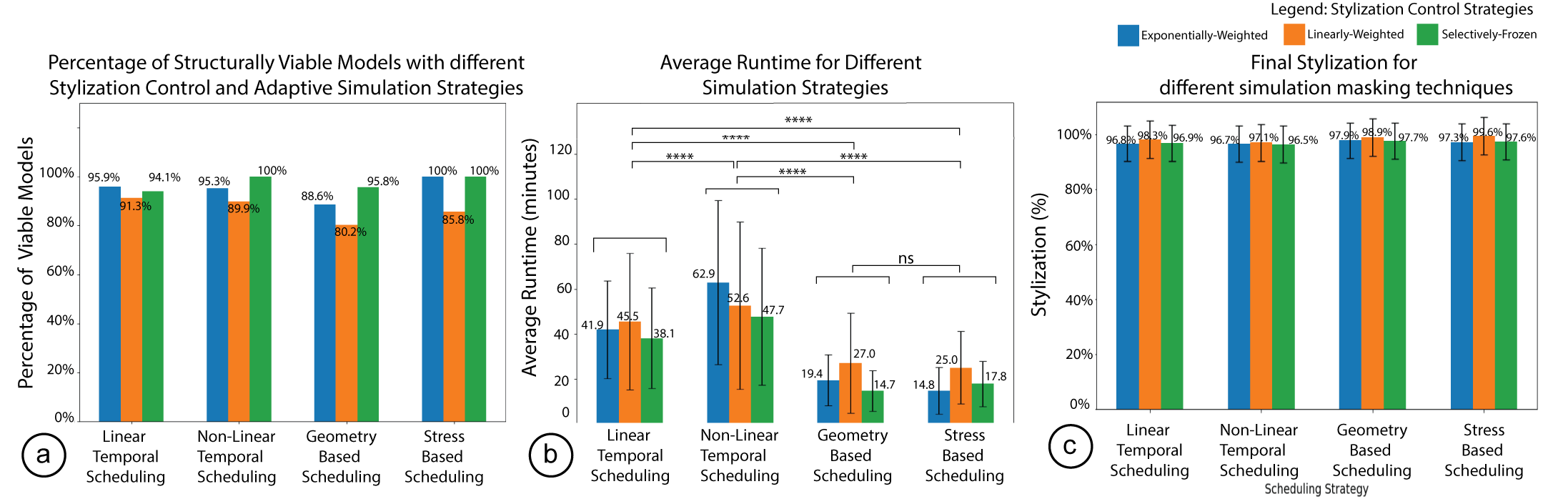}
    \caption{Technical Evaluation: Using our dataset of 30 models, we evaluate a) what percentage of models were structurally viable after stylization between all possible combinations of simulation control strategies and adaptive simulation strategies, b) the average runtime of the different simulation strategies, c) the average style-loss between the different stylization control strategies compared to the baseline of original (freestyle) stylization. 'Stress-based' scheduling with `exponentially-weighted' stylization, and `Stress-based' scheduling with `selectively-frozen' stylization provide the best trade-off between structural viability, run-time and style-loss.  (*p < 0.05, **p < 0.01, ***p < 0.001, ****p < 0.0001)}
    \vspace{-3mm}
    \label{fig:technical_eval}
\end{figure*}

\subsection{Evaluating Structural Viability of Models}
We evaluated the structural viability of the stylized models under different MechStyle configurations using the same approach as in the formative study. While in the formative study, only 25.55\% of models were structurally viable after stylization, MechStyle improved the result in all configurations with the best result being 100\% of models being structurally viable and the worst-case being 80.2\% of models being structurally viable (Figure ~\ref{fig:technical_eval}a). Three configurations resulted in all models being structurally viable: (1)~Non-Linear Temporal Scheduling with selectively frozen stylization control, (2)~Stress-based scheduling with exponentially weighted stylization control, and (3)~Stress-based scheduling with selectively frozen stylization control. Thus, these three combinations of techniques are robust in identifying structural issues and preventing the model from becoming unviable during stylization. 

\subsection{Evaluating Run Time with Adaptive Scheduling Strategies}

To evaluate which MechStyle configuration is the most run-time efficient, we extracted the average run-times for each combination of stylization control strategy and scheduling strategy as shown in Figure~\ref{fig:technical_eval}b. Since the runtime is affected by simulations, we first compare simulation strategies. Between the simulation strategies, a Welch’s ANOVA test showed a significant effect~($F = 452.72$, $p~<~0.0001$) in the system run-time. A Games Howell post-hoc analysis shows that the `Non-Linear Scheduling' strategy increased the system's runtime ($T = 854.73$, $p~<~0.0001$). However, the `Geometry-Based' ($T = 16.03$, $p~<~0.0001$) and `Stress-Based' ($T = 18.02$, $p~<~0.0001$) adaptive scheduling strategies significantly reduce the overall runtime of stylization as compared to the control. There was no significant difference ($p~>~0.1$) between `Geometry-Based' and `Stress-Based' conditions. Next, we compare the six different combinations of the three stylization control strategies and these two scheduling techniques. A Welch’s ANOVA test showed a significant effect~($F = 26.20$, $p~<~0.0001$) in the system run-time between these 6 conditions.  A Games Howell post-hoc analysis showed that for the `Geometry-based' scheduling technique, the `selectively-frozen' stylization control strategy was significantly more efficient than Linearly-weighed ($T = 6.54$, $p~<~0.0001$) and Exponentially-weighed ($T = 4.19$, $p~<~0.0001$). For `Stress-based' scheduling, the `exponentially-weighted' stylization control strategy was significantly more efficient than the Linearly-weighed strategy ($T = 6.74$, $p~<~0.0001$). However, we found no statistically significant difference between `exponentially-weighted' and `selectively-frozen' ($T = 6.74$, $p~>~0.05$). When comparing these two with the configuration of `Geometry-based' scheduling technique with `selectively-frozen' stylization control we found no statistically significant difference ($p~>~0.05$). 

\subsection{Evaluating Style Loss with Stylization Control Strategies}
\faraz{MechStyle preserves the structural integrity of 3D models by reducing stylization in local areas, which affects the final stylization result. Prior work on stylization~\cite{x_mesh, michel2022text2mesh} use `style-loss' as a metric used to evaluate the quality of stylization. Style loss is described as a semantic loss that measures the similarity between a target text description and the generated stylized mesh. It is based on the multi-modal embedding space provided by the pre-trained CLIP~\cite{radford2021learning} model. The semantic loss compares CLIP embedding of rendered views of the stylized mesh with embeddings of the target text. Since MechStyle effectively controls stylization to reduce its impact on the structural properties of the model, we assume that the model stylized without MechStyle attains convergence on the style. This allows to evaluate how MechStyle's influence on the stylization capability of the Generative AI model.  Thus, we compare the style-loss of the MechStyle result with a freely stylized model, representing full stylization. We record the stylization loss for all configurations as a percentage of that of the fully stylized model.} Figure~\ref{fig:technical_eval}c shows the stylization quality of all different techniques. A Welch’s ANOVA test did not show any significant difference ($p > 0.05$) in the stylization loss, showing that none of the strategies affected the overall stylization capability.

\subsection{Summary of Findings}


Our goal in this technical evaluation was to evaluate our stylization control strategies and adaptive scheduling strategies based on the trade-off between the objectives of structural viability of stylized models, the total run-time, and the final stylization quality achieved. For this purpose, we conducted a grid search over all possible combinations of adaptive scheduling techniques and stylization control strategies. For structural integrity~(\autoref{fig:technical_eval}a), we find that three combinations: (1)~`Non-Linear Temporal' scheduling with `selectively frozen` stylization, (2)~`Stress-based' scheduling with `exponentially weighted' stylization, and (3)~`Stress-based' scheduling with `selectively frozen' stylization, resulted in all stylized models being structurally viable. For overall runtime, we find that the configurations of (1) `Geometry-based' scheduling with `selectively-frozen', (2) `Stress-based' scheduling with `exponentially-weighted' stylization, and (3) `Stress-based' scheduling with `selectively-frozen' stylization gave the most significantly efficient runtime. Finally, we found no statistically significant loss in the stylization achieved by any of the configurations.

The approaches that perform  best across all three objectives are (1) 'Stress-based' scheduling with `exponentially-weighted' stylization, and (2) `Stress-based' scheduling with `selectively-frozen' stylization. Thus, we conclude that these two strategies are most effective to use in MechStyle. 


%% file: Sections/07_implementation.tex
\section{Implementation}
\faraz{In this section, we describe the infrastructure we used for our experiments and the technical details of our stylization and simulation environments. Given the high number of configurations in our technical evaluation, we conducted our experiments on a single machine with NVIDIA-L4 GPU. However, MechStyle can run on any CUDA-based GPU with 8 GB GPU memory. The tool would also work with a system without a GPU, however execution speeds would be affected due to the absence of GPU processing. We use the stylization method from X-Mesh~\cite{x_mesh} written in Pytorch and conduct simulations using the WARP~\cite{warp2022} library. Since stylization is performed on triangular surface meshes while simulations are performed on volumetric meshes (tetrahedral), we developed a method to convert intermediate stylized models into volumetric representations with fTetWild~\cite{hu2020fast}. }

Stylization techniques primarily work on surface meshes (triangular meshes, particularly OBJ files) because surface meshes are efficient to store and manipulate the vertices and colors. However, mechanical simulation requires a volumetric tetrahedral representation (TET) to accurately capture the internal structure and dynamics of objects. Since we require simulation results (per-element stress values) to inform the stylization module, it is important to find a mapping between the OBJ and TET representations. For simulation, we convert the OBJ model into a TET model using fTetWild~\cite{hu2020fast}. We find a mapping between the TET representation and the OBJ representation by normalizing the sizes of the two models to a unit bounding box, extracting the external surface of the TET model, and running K Nearest Neighbors $(k=1)$ to find the closest matching points between the two representations. Finally, we compute the local thickness of a 3D model using the Shrinking Sphere method~\cite{inui2016shrinking} due to its robustness and computation speed. 

We will open-source the codebase for MechStyle with documentation to support future work in this area. 



%% file: Sections/08_applications.tex
\section{Applications}
\faraz{In this section, we present examples of 3D models stylized with MechStyle that demonstrate the value of stylizing 3D models while taking its structural integrity into account. We demonstrate four different application scenarios across three categories: Personal Accessories, Home Decor, and Personalized Assistive Devices. For each model, we demonstrate the `Freestyle' stylization result (without MechStyle), and its FEA simulation showing the impact on its structural properties. Next, we show the result generated with MechStyle, along with its structural simulation, showing how MechStyle generates results with similar style changes while preserving its structural properties. We printed our models with a Stratasys J55 printer and Vero material. We updated MechStyle with the corresponding material properties.}

\begin{figure*}[!ht]
    \centering
    \includegraphics[width=0.8\textwidth]{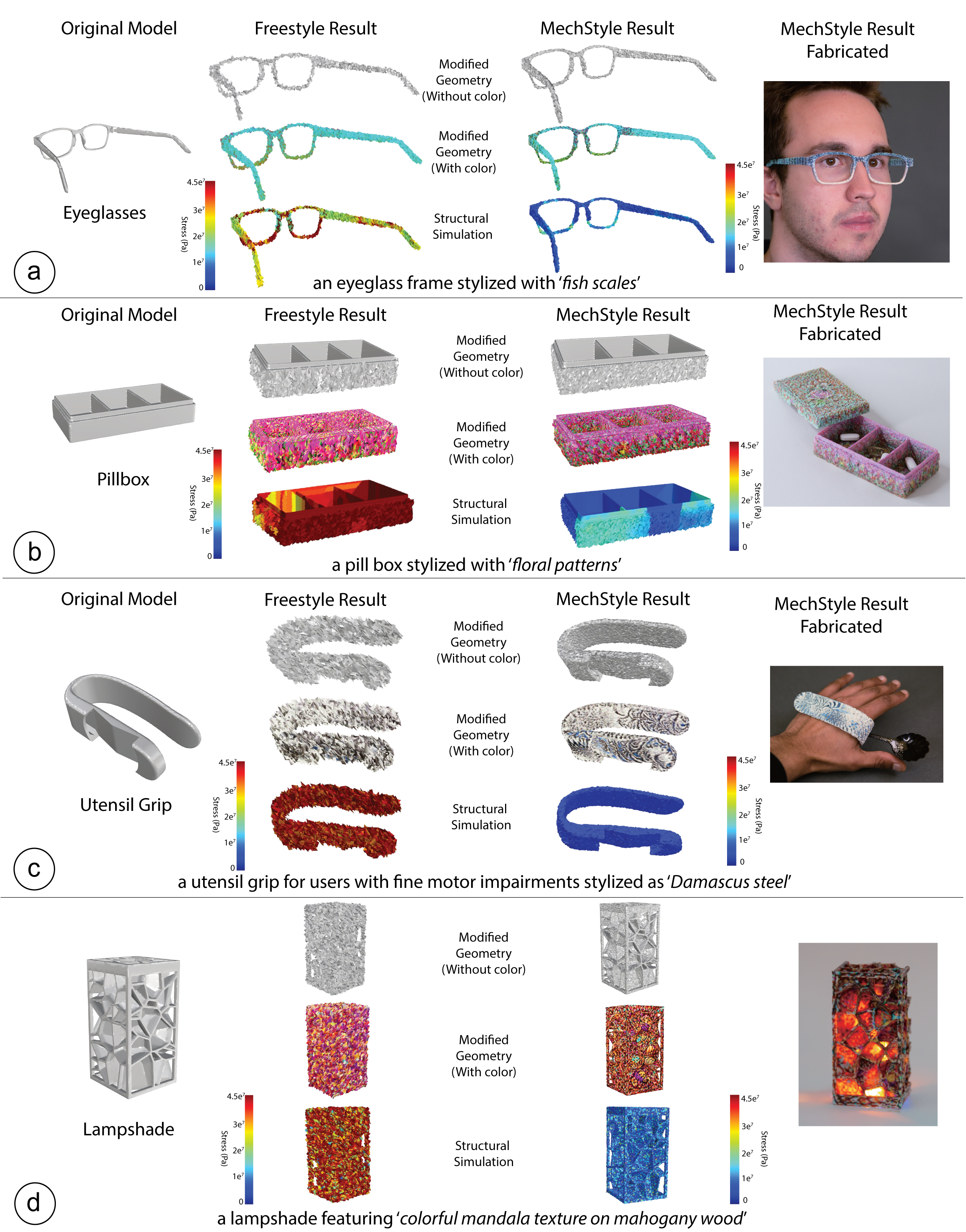}
  \caption{\faraz{MechStyle enables creators to stylize 3D models with text prompts while preserving their structural integrity. Here, we used MechStyle to stylize five 3D models while ensuring that the printed objects do not break when accidentally dropped by the user. For each model, we demonstrate the Freestyle (without MechStyle) result, and the MechStyle result, showing how using MechStyle allows users to stylize 3D models while retaining its structural integrity: a)~an eye-glass frame stylized with `\textit{fish scales}'. b)~a pill box with \textit{floral patterns}', c)~a utensil grip for users with fine motor impairments stylized as `\textit{Damascus steel}' pattern of the user's silverware set, and e)~a lampshade featuring `\textit{colorful mandala texture on mahogany wood}'.}}

    \label{fig:applications}
\end{figure*}


\subsection{Personalized Accessories}

Fabricating personalized accessories is one of the applications for consumer fabrication. Many novice makers desire the flexibility to create personalized and expressive objects without having the expertise to design and stylize these models without AI assistance~\cite{faruqi2023style2fab}. MechStyle allows users to stylize 3D models while maintaining their structural integrity under the forces applied on the model when dropped. Here, we demonstrate an eyeglass 3D model stylized with MechStyle to have a ``\textit{fish scale}'' texture. The glass frame model we selected has multiple fragile segments, such as the thin bridge of the nose (\autoref{fig:applications}a). \faraz{Here, we show both the Freestyle stylization result, and the MechStyle result, along with an FEA simulation of the two stylized 3D models. The Freestyle stylization modified the geometry of the eyeglasses such that several segments in the original model were significantly weakened, such as the bridge and the frame segments. On the other hand, MechStyle modified the geometry while preserving the structurally weak segments such that the geometrical changes do not introduce structural issues. Thus, the FEA simulation of the MechStyle result has no significantly weaker segments, and can be fabricated and used.} 

As another example, we stylize a pill box to have the style of ``\textit{floral patterns}'' on the outside of the box (using Style2Fab~\cite{faruqi2023style2fab} to selectively only stylize the outside) (\autoref{fig:applications}b). \faraz{On comparing the Freestyle and MechStyle result, we see that both methods modified the outer geometry of the model while retaining the inner geometry. However, the FEA simulation result of the MechStyle result shows structural viability, while the Freestyle result has several regions that are significantly weakened by the stylization. MechStyle preserves the model's structural integrity in regions such as the thin edges of the box and lid so that the model can survive accidental drops, thus preserving its structural integrity for fabrication.}


\subsection{Personalized Assistive Technologies}

Like personalized accessories, digital fabrication is a popular approach for creating personalized assistive technologies. Unlike general-purpose accessories for which users have a variety of off-the-shelf styles available, assistive devices are rarely personalized and frequently have a stigmatizing medical appearance. Social Accessibility is a critical topic of accessibility research that explores how the appearance of assistive technologies affects the stigma placed on people with disabilities and how people with disabilities stylize their devices to express a positive disabled identity~\cite{Shinohara_social}. Many people have used digital fabrication tools to stylize their assistive technology, such as decorative hearing aids and cochlear implants~\cite{Profita_hearing} or 3D printing discrete hand splints \cite{Hofmann_OT}. \faraz{Here, we demonstrate personalization of a utensil grip for people with fine motor impairments to match the ``\textit{Damascus steel}'' pattern of the user's decorative silverware set. \autoref{fig:applications}c, shows the stylization result of the utensil grip, both with Freestyle and MechStyle stylization. Here, we observe that Freestyle modified the geometry significantly more than MechStyle, and its FEA simulation shows that most of the region has been significantly weakened, with the stress values being above the fabrication material's yield stress value. In contrast, MechStyle incorporated intermediate simulation results during the stylization process, and thus the stylized result has lesser geometrical changes, which results in a structurally viable, stylized result. Thus, the stylized model has the desired style and is structurally robust enough to survive accidental drops. }


\subsection{Home Decor}

Similar to personal accessories, interior decoration is a popular domain for consumer fabrication with models that have both aesthetic and structural requirements. In this application, we stylized a lampshade in the style of ``\textit{colorful mandala texture on mahogany wood}'' shown in \autoref{fig:applications}e. The lampshade has intricate structural panels that have differently angled surfaces, creating a dynamic lighting effect from a light bulb placed inside. \faraz{We demonstrate both Freestyle and MechStyle's result. Here, intricate panels on the lampshade require careful stylization to preserve their structural integrity. We observe that the Freestyle stylization modified the 3D geometry such that even though the style is accurately demonstrated, the stylized result is significantly weakened as shown by the FEA simulation. In contrast, MechStyle's controlled the modification of the 3D geometry such that it doesn't impact the structural integrity of the model. Thus, MechStyle's result shows a modified 3D model with its FEA simulation showing no significantly weak sections. Meanwhile, the color and geometrical variations in different segments amplify the lighting effect of the lampshade. }


%% file: Sections/09_discussion.tex
\section{Discussion and Limitations}

In developing MechStyle, we make three assumptions to find a tractable solution that incorporates simulation in a Generative AI stylization process. Each of these assumptions limits the capabilities of the system and reflects opportunities for future work. First, we assume that modelers provide a structurally valid model that satisfies all of its requirements in the real world. Second, we assume a direct correlation between simulated stress and the mechanics of a model post-fabrication. Third, we assume that a drop test is a sufficient approximation of the stresses put on a model in regular use. Our evaluation and demonstrations show that these assumptions hold under a variety of scenarios. However, MechStyle can be improved in several ways.

\subsection{Structural Validity of Open Source Models}
MechStyle works by preserving the structural integrity of the original model. If that original model is structurally valid, MechStyle will ensure its validity during stylization. However, if a user inputs a 3D model that is not structurally valid, MechStyle cannot ensure structural validity. Thus, if the user loads an open-source model that is structurally unviable, MechStyle will communicate this as a warning with the FEA view showing the regions of high stress. 

However, a modified version of MechStyle can be used to improve the structural integrity of structurally unviable 3D models. In its current form, MechStyle assumes that the stress in the model is under the material's critical value indicating structural viability, and that stylization should be controlled if the stress increases. However, stylization can also be used to decrease the local stresses by guiding it to thicken the local regions. Thus, a future system could modify MechStyle by introducing a novel stylization control strategy designed for structurally unviable 3D models. This new strategy could identify and reduce the local stress to be under the critical value of the material. 



\subsection{Towards Holistic and Generalizable Mechanical Simulation}
We assume that a mechanical simulation, such as Finite Element Analysis (FEA), will reveal the parts of a model that must be preserved to ensure it is structurally viable after stylization. However, these simulations only account for a subset of all real-world constraints placed on a model. For instance, Finite Element Analysis relies only on model geometry and a specified model material. However, other structural failures can result from complex relationships between geometry, fabrication method, slicer settings, and usage scenarios. In a traditional CAD pipeline, an expert user may be able to specify each of these features. However, in a system such as MechStyle, all these constraints would have to be automatically captured. Alternatively, we could imagine a system that simulates these real-world conditions and materials but this poses several non-trivial challenges, including the known gap between simulated and real-world behaviors (sim-to-real gap~\cite{wang2024diffusebot}). Since MechStyle does not rely on the specific implementation of mechanical simulation, improved and customized simulations can be added to the system.

\faraz{Secondly, MechStyle uses a model- and use case-agnostic mechanical simulation, specifically a `drop-test', to assess structural integrity. Such drop-tests has been demonstrated as an effective generalizable method in prior work on stress analysis~\cite{stress_relief, stochastic_structural_analysis} to identify weak regions in the 3D model}. However, this method does not evaluate the model's structural integrity given specific usage scenarios. In a traditional CAD pipeline, an expert will manually specify the forces and their locations for the different usage scenarios for an accurate FEA simulation. However, since our target users are non-experts who may not be able to identify the load requirements of a model apriori, we cannot rely on these inputs in MechStyle. \faraz{Future work in this topic can focus on making this specification task accessible to novices by developing new methods to extract implicit constraints from usage data. One approach can be modeling applied forces and their locations based on a video of the modeler using the original model for a specific activity.}


\faraz{
\subsection{Visual Plausibility of Stylization}
The visual plausibility of textures generated by stylization plays a critical role in usability and adoption of these these stylization methods. Studies like those conducted in Text2Mesh~\cite{michel2022text2mesh} and Style2Fab~\cite{faruqi2023style2fab}, evaluate the user perception of visual quality, and style fidelity, highlight the importance of assessing how well stylizations align with both user expectations. In MechStyle, our goal was to develop a tool that extends such stylization methods to create models that are not only visually appealing but also structurally viable. However, some styles generated through MechStyle and related tools may lack realism. This can arise from several factors, such as limitations in the training data used for stylization models resulting in hallucinated patterns that deviate from real-world appearances, or poor mesh quality that prevents the style from aligning properly with the underlying geometry. As Generative AI methods continue to evolve, newer and more advanced stylization techniques are being developed, offering increasingly realistic and diverse styles. MechStyle’s framework is designed to integrate seamlessly with any geometry-based stylization method, leveraging their advancements while ensuring structural viability.}

%% file: Sections/10_conclusion.tex
\section{Conclusion}
In this paper, we proposed a new approach to the personalization of 3D models that allows users to stylize 3D models with Generative AI while ensuring they retain their structural integrity. A formative study of 3D models sourced from Thingiverse showed that Generative AI-based stylization impacts the structural integrity of 3D models, thus necessitating to incorporate mechanical simulation into the iterative stylization process. We presented an approach that combines stylization and mechanical simulation. We then studied different stylization control strategies and adaptive scheduling strategies for the simulation to maximize stylization outcomes while preserving structural integrity. Our technical evaluation shows that none of the stylization control strategies and adaptive scheduling strategies significantly impacted the stylization capability of the system and that 'Stress-based' scheduling with `exponentially-weighted' stylization, and `Stress-based' scheduling with `selectively-frozen' stylization are the most efficient in terms of runtime and generating structurally viable 3D models. 

%% file: main.bbl

\begin{thebibliography}{49}


\ifx \showCODEN    \undefined \def \showCODEN     #1{\unskip}     \fi
\ifx \showISBNx    \undefined \def \showISBNx     #1{\unskip}     \fi
\ifx \showISBNxiii \undefined \def \showISBNxiii  #1{\unskip}     \fi
\ifx \showISSN     \undefined \def \showISSN      #1{\unskip}     \fi
\ifx \showLCCN     \undefined \def \showLCCN      #1{\unskip}     \fi
\ifx \shownote     \undefined \def \shownote      #1{#1}          \fi
\ifx \showarticletitle \undefined \def \showarticletitle #1{#1}   \fi
\ifx \showURL      \undefined \def \showURL       {\relax}        \fi
\providecommand\bibfield[2]{#2}
\providecommand\bibinfo[2]{#2}
\providecommand\natexlab[1]{#1}
\providecommand\showeprint[2][]{arXiv:#2}

\bibitem[Abdullah et~al\mbox{.}(2021)]%
        {abdullah2021fastforce}
\bibfield{author}{\bibinfo{person}{Muhammad Abdullah}, \bibinfo{person}{Martin Taraz}, \bibinfo{person}{Yannis Kommana}, \bibinfo{person}{Shohei Katakura}, \bibinfo{person}{Robert Kovacs}, \bibinfo{person}{Jotaro Shigeyama}, \bibinfo{person}{Thijs Roumen}, {and} \bibinfo{person}{Patrick Baudisch}.} \bibinfo{year}{2021}\natexlab{}.
\newblock \showarticletitle{FastForce: Real-Time Reinforcement of Laser-Cut Structures}. In \bibinfo{booktitle}{\emph{Proceedings of the 2021 CHI Conference on Human Factors in Computing Systems}}. \bibinfo{pages}{1--12}.
\newblock


\bibitem[Alcock et~al\mbox{.}(2016)]%
        {alcock2016barriers}
\bibfield{author}{\bibinfo{person}{Celena Alcock}, \bibinfo{person}{Nathaniel Hudson}, {and} \bibinfo{person}{Parmit~K. Chilana}.} \bibinfo{year}{2016}\natexlab{}.
\newblock \showarticletitle{Barriers to Using, Customizing, and Printing 3D Designs on Thingiverse}. In \bibinfo{booktitle}{\emph{Proceedings of the 2016 ACM International Conference on Supporting Group Work}} (Sanibel Island, Florida, USA) \emph{(\bibinfo{series}{GROUP '16})}. \bibinfo{publisher}{Association for Computing Machinery}, \bibinfo{address}{New York, NY, USA}, \bibinfo{pages}{195–199}.
\newblock
\showISBNx{9781450342766}
\href{https://doi.org/10.1145/2957276.2957301}{doi:\nolinkurl{10.1145/2957276.2957301}}


\bibitem[Berman et~al\mbox{.}(2021)]%
        {berman2021howdiy}
\bibfield{author}{\bibinfo{person}{Alexander Berman}, \bibinfo{person}{Ketan Thakare}, \bibinfo{person}{Joshua Howell}, \bibinfo{person}{Francis Quek}, {and} \bibinfo{person}{Jeeeun Kim}.} \bibinfo{year}{2021}\natexlab{}.
\newblock \showarticletitle{HowDIY: Towards Meta-Design Tools to Support Anyone to 3D Print Anywhere}. In \bibinfo{booktitle}{\emph{26th International Conference on Intelligent User Interfaces}} (College Station, TX, USA) \emph{(\bibinfo{series}{IUI '21})}. \bibinfo{publisher}{Association for Computing Machinery}, \bibinfo{address}{New York, NY, USA}, \bibinfo{pages}{491–503}.
\newblock
\showISBNx{9781450380171}
\href{https://doi.org/10.1145/3397481.3450638}{doi:\nolinkurl{10.1145/3397481.3450638}}


\bibitem[Chaudhuri et~al\mbox{.}(2013)]%
        {chaudhuri2013attribit}
\bibfield{author}{\bibinfo{person}{Siddhartha Chaudhuri}, \bibinfo{person}{Evangelos Kalogerakis}, \bibinfo{person}{Stephen Giguere}, {and} \bibinfo{person}{Thomas Funkhouser}.} \bibinfo{year}{2013}\natexlab{}.
\newblock \showarticletitle{Attribit: Content Creation with Semantic Attributes}. In \bibinfo{booktitle}{\emph{Proceedings of the 26th Annual ACM Symposium on User Interface Software and Technology}} (St. Andrews, Scotland, United Kingdom) \emph{(\bibinfo{series}{UIST '13})}. \bibinfo{publisher}{Association for Computing Machinery}, \bibinfo{address}{New York, NY, USA}, \bibinfo{pages}{193–202}.
\newblock
\showISBNx{9781450322683}
\href{https://doi.org/10.1145/2501988.2502008}{doi:\nolinkurl{10.1145/2501988.2502008}}


\bibitem[Chowdhery et~al\mbox{.}(2023)]%
        {chowdhery2023palm}
\bibfield{author}{\bibinfo{person}{Aakanksha Chowdhery}, \bibinfo{person}{Sharan Narang}, \bibinfo{person}{Jacob Devlin}, \bibinfo{person}{Maarten Bosma}, \bibinfo{person}{Gaurav Mishra}, \bibinfo{person}{Adam Roberts}, \bibinfo{person}{Paul Barham}, \bibinfo{person}{Hyung~Won Chung}, \bibinfo{person}{Charles Sutton}, \bibinfo{person}{Sebastian Gehrmann}, {et~al\mbox{.}}} \bibinfo{year}{2023}\natexlab{}.
\newblock \showarticletitle{Palm: Scaling language modeling with pathways}.
\newblock \bibinfo{journal}{\emph{Journal of Machine Learning Research}} \bibinfo{volume}{24}, \bibinfo{number}{240} (\bibinfo{year}{2023}), \bibinfo{pages}{1--113}.
\newblock


\bibitem[com(2023)]%
        {Thingiverse2023Apr}
\bibfield{author}{\bibinfo{person}{Thingiverse. com}.} \bibinfo{year}{2023}\natexlab{}.
\newblock \bibinfo{title}{{Thingiverse - Digital Designs for Physical Objects}}.
\newblock
\urldef\tempurl%
\url{https://www.thingiverse.com}
\showURL{%
\tempurl}
\newblock
\shownote{[Online; accessed 4. Apr. 2023]}.


\bibitem[Community(2018)]%
        {blenderUI}
\bibfield{author}{\bibinfo{person}{Blender~Online Community}.} \bibinfo{year}{2018}\natexlab{}.
\newblock \bibinfo{booktitle}{\emph{Blender - a 3D modelling and rendering package}}.
\newblock Blender Foundation, Stichting Blender Foundation, Amsterdam.
\newblock
\urldef\tempurl%
\url{http://www.blender.org}
\showURL{%
\tempurl}


\bibitem[Decatur et~al\mbox{.}(2023)]%
        {decatur20233d}
\bibfield{author}{\bibinfo{person}{Dale Decatur}, \bibinfo{person}{Itai Lang}, {and} \bibinfo{person}{Rana Hanocka}.} \bibinfo{year}{2023}\natexlab{}.
\newblock \showarticletitle{3d highlighter: Localizing regions on 3d shapes via text descriptions}. In \bibinfo{booktitle}{\emph{Proceedings of the IEEE/CVF Conference on Computer Vision and Pattern Recognition}}. \bibinfo{pages}{20930--20939}.
\newblock


\bibitem[Faruqi et~al\mbox{.}(2021)]%
        {faruqi2021slicehub}
\bibfield{author}{\bibinfo{person}{Faraz Faruqi}, \bibinfo{person}{Kenneth Friedman}, \bibinfo{person}{Leon Cheng}, \bibinfo{person}{Michael Wessely}, \bibinfo{person}{Sriram Subramanian}, {and} \bibinfo{person}{Stefanie Mueller}.} \bibinfo{year}{2021}\natexlab{}.
\newblock \showarticletitle{SliceHub: Augmenting Shared 3D Model Repositories with Slicing Results for 3D Printing}.
\newblock \bibinfo{journal}{\emph{arXiv preprint arXiv:2109.14722}} (\bibinfo{year}{2021}).
\newblock


\bibitem[Faruqi et~al\mbox{.}(2023)]%
        {faruqi2023style2fab}
\bibfield{author}{\bibinfo{person}{Faraz Faruqi}, \bibinfo{person}{Ahmed Katary}, \bibinfo{person}{Tarik Hasic}, \bibinfo{person}{Amira Abdel-Rahman}, \bibinfo{person}{Nayeemur Rahman}, \bibinfo{person}{Leandra Tejedor}, \bibinfo{person}{Mackenzie Leake}, \bibinfo{person}{Megan Hofmann}, {and} \bibinfo{person}{Stefanie Mueller}.} \bibinfo{year}{2023}\natexlab{}.
\newblock \showarticletitle{Style2Fab: Functionality-Aware Segmentation for Fabricating Personalized 3D Models with Generative AI}. In \bibinfo{booktitle}{\emph{Proceedings of the 36th Annual ACM Symposium on User Interface Software and Technology}}. \bibinfo{pages}{1--13}.
\newblock


\bibitem[Faruqi et~al\mbox{.}(2025)]%
        {faruqi2025tactstyle}
\bibfield{author}{\bibinfo{person}{Faraz Faruqi}, \bibinfo{person}{Maxine Perroni-Scharf}, \bibinfo{person}{Jaskaran~Singh Walia}, \bibinfo{person}{Yunyi Zhu}, \bibinfo{person}{Shuyue Feng}, \bibinfo{person}{Donald Degraen}, {and} \bibinfo{person}{Stefanie Mueller}.} \bibinfo{year}{2025}\natexlab{}.
\newblock \showarticletitle{TactStyle: Generating Tactile Textures with Generative AI for Digital Fabrication}. In \bibinfo{booktitle}{\emph{Proceedings of the 2025 CHI Conference on Human Factors in Computing Systems}}. \bibinfo{pages}{1--16}.
\newblock


\bibitem[Faruqi et~al\mbox{.}(2024)]%
        {faruqi2024shaping}
\bibfield{author}{\bibinfo{person}{Faraz Faruqi}, \bibinfo{person}{Yingtao Tian}, \bibinfo{person}{Vrushank Phadnis}, \bibinfo{person}{Varun Jampani}, {and} \bibinfo{person}{Stefanie Mueller}.} \bibinfo{year}{2024}\natexlab{}.
\newblock \showarticletitle{Shaping realities: Enhancing 3D generative AI with fabrication constraints}.
\newblock \bibinfo{journal}{\emph{arXiv preprint arXiv:2404.10142}} (\bibinfo{year}{2024}).
\newblock


\bibitem[Gao et~al\mbox{.}(2022)]%
        {gao2022get3d}
\bibfield{author}{\bibinfo{person}{Jun Gao}, \bibinfo{person}{Tianchang Shen}, \bibinfo{person}{Zian Wang}, \bibinfo{person}{Wenzheng Chen}, \bibinfo{person}{Kangxue Yin}, \bibinfo{person}{Daiqing Li}, \bibinfo{person}{Or Litany}, \bibinfo{person}{Zan Gojcic}, {and} \bibinfo{person}{Sanja Fidler}.} \bibinfo{year}{2022}\natexlab{}.
\newblock \showarticletitle{GET3D: A Generative Model of High Quality 3D Textured Shapes Learned from Images}. In \bibinfo{booktitle}{\emph{Advances In Neural Information Processing Systems}}.
\newblock


\bibitem[Gao et~al\mbox{.}(2023)]%
        {gao2023textdeformer}
\bibfield{author}{\bibinfo{person}{William Gao}, \bibinfo{person}{Noam Aigerman}, \bibinfo{person}{Thibault Groueix}, \bibinfo{person}{Vladimir~G Kim}, {and} \bibinfo{person}{Rana Hanocka}.} \bibinfo{year}{2023}\natexlab{}.
\newblock \showarticletitle{TextDeformer: Geometry Manipulation using Text Guidance}.
\newblock \bibinfo{journal}{\emph{arXiv preprint arXiv:2304.13348}} (\bibinfo{year}{2023}).
\newblock


\bibitem[Hofmann et~al\mbox{.}(2018)]%
        {hofmann2018greater}
\bibfield{author}{\bibinfo{person}{Megan Hofmann}, \bibinfo{person}{Gabriella Hann}, \bibinfo{person}{Scott~E. Hudson}, {and} \bibinfo{person}{Jennifer Mankoff}.} \bibinfo{year}{2018}\natexlab{}.
\newblock \showarticletitle{Greater than the Sum of Its PARTs: Expressing and Reusing Design Intent in 3D Models}. In \bibinfo{booktitle}{\emph{Proceedings of the 2018 CHI Conference on Human Factors in Computing Systems}} (Montreal QC, Canada) \emph{(\bibinfo{series}{CHI '18})}. \bibinfo{publisher}{Association for Computing Machinery}, \bibinfo{address}{New York, NY, USA}, \bibinfo{pages}{1–12}.
\newblock
\showISBNx{9781450356206}
\href{https://doi.org/10.1145/3173574.3173875}{doi:\nolinkurl{10.1145/3173574.3173875}}


\bibitem[Hofmann et~al\mbox{.}(2019)]%
        {Hofmann_OT}
\bibfield{author}{\bibinfo{person}{Megan Hofmann}, \bibinfo{person}{Kristin Williams}, \bibinfo{person}{Toni Kaplan}, \bibinfo{person}{Stephanie Valencia}, \bibinfo{person}{Gabriella Hann}, \bibinfo{person}{Scott~E. Hudson}, \bibinfo{person}{Jennifer Mankoff}, {and} \bibinfo{person}{Patrick Carrington}.} \bibinfo{year}{2019}\natexlab{}.
\newblock \showarticletitle{"Occupational Therapy is Making": Clinical Rapid Prototyping and Digital Fabrication}. In \bibinfo{booktitle}{\emph{Proceedings of the 2019 CHI Conference on Human Factors in Computing Systems}} (Glasgow, Scotland Uk) \emph{(\bibinfo{series}{CHI '19})}. \bibinfo{publisher}{Association for Computing Machinery}, \bibinfo{address}{New York, NY, USA}, \bibinfo{pages}{1–13}.
\newblock
\showISBNx{9781450359702}
\href{https://doi.org/10.1145/3290605.3300544}{doi:\nolinkurl{10.1145/3290605.3300544}}


\bibitem[Hu et~al\mbox{.}(2020)]%
        {hu2020fast}
\bibfield{author}{\bibinfo{person}{Yixin Hu}, \bibinfo{person}{Teseo Schneider}, \bibinfo{person}{Bolun Wang}, \bibinfo{person}{Denis Zorin}, {and} \bibinfo{person}{Daniele Panozzo}.} \bibinfo{year}{2020}\natexlab{}.
\newblock \showarticletitle{Fast tetrahedral meshing in the wild}.
\newblock  (\bibinfo{year}{2020}).
\newblock


\bibitem[Hudson et~al\mbox{.}(2016)]%
        {hudson2016understanding}
\bibfield{author}{\bibinfo{person}{Nathaniel Hudson}, \bibinfo{person}{Celena Alcock}, {and} \bibinfo{person}{Parmit~K. Chilana}.} \bibinfo{year}{2016}\natexlab{}.
\newblock \showarticletitle{Understanding Newcomers to 3D Printing: Motivations, Workflows, and Barriers of Casual Makers}. In \bibinfo{booktitle}{\emph{Proceedings of the 2016 CHI Conference on Human Factors in Computing Systems}} (San Jose, California, USA) \emph{(\bibinfo{series}{CHI '16})}. \bibinfo{publisher}{Association for Computing Machinery}, \bibinfo{address}{New York, NY, USA}, \bibinfo{pages}{384–396}.
\newblock
\showISBNx{9781450333627}
\href{https://doi.org/10.1145/2858036.2858266}{doi:\nolinkurl{10.1145/2858036.2858266}}


\bibitem[Inui et~al\mbox{.}(2016)]%
        {inui2016shrinking}
\bibfield{author}{\bibinfo{person}{Masatomo Inui}, \bibinfo{person}{Nobuyuki Umezu}, {and} \bibinfo{person}{Ryohei Shimane}.} \bibinfo{year}{2016}\natexlab{}.
\newblock \showarticletitle{Shrinking sphere: A parallel algorithm for computing the thickness of 3D objects}.
\newblock \bibinfo{journal}{\emph{Computer-Aided Design and Applications}} \bibinfo{volume}{13}, \bibinfo{number}{2} (\bibinfo{year}{2016}), \bibinfo{pages}{199--207}.
\newblock


\bibitem[Johns et~al\mbox{.}(2023)]%
        {multiple_objectives_uist_23}
\bibfield{author}{\bibinfo{person}{Christoph~Albert Johns}, \bibinfo{person}{Jo\~{a}o~Marcelo Evangelista~Belo}, \bibinfo{person}{Anna~Maria Feit}, \bibinfo{person}{Clemens~Nylandsted Klokmose}, {and} \bibinfo{person}{Ken Pfeuffer}.} \bibinfo{year}{2023}\natexlab{}.
\newblock \showarticletitle{Towards Flexible and Robust User Interface Adaptations With Multiple Objectives} \emph{(\bibinfo{series}{UIST '23})}. \bibinfo{publisher}{Association for Computing Machinery}, \bibinfo{address}{New York, NY, USA}, Article \bibinfo{articleno}{108}, \bibinfo{numpages}{17}~pages.
\newblock
\showISBNx{9798400701320}
\href{https://doi.org/10.1145/3586183.3606799}{doi:\nolinkurl{10.1145/3586183.3606799}}


\bibitem[Koyama et~al\mbox{.}(2015)]%
        {koyama2015autoconnect}
\bibfield{author}{\bibinfo{person}{Yuki Koyama}, \bibinfo{person}{Shinjiro Sueda}, \bibinfo{person}{Emma Steinhardt}, \bibinfo{person}{Takeo Igarashi}, \bibinfo{person}{Ariel Shamir}, {and} \bibinfo{person}{Wojciech Matusik}.} \bibinfo{year}{2015}\natexlab{}.
\newblock \showarticletitle{AutoConnect: Computational Design of 3D-Printable Connectors}.
\newblock \bibinfo{journal}{\emph{ACM Trans. Graph.}} \bibinfo{volume}{34}, \bibinfo{number}{6}, Article \bibinfo{articleno}{231} (\bibinfo{date}{nov} \bibinfo{year}{2015}), \bibinfo{numpages}{11}~pages.
\newblock
\showISSN{0730-0301}
\href{https://doi.org/10.1145/2816795.2818060}{doi:\nolinkurl{10.1145/2816795.2818060}}


\bibitem[Kuznetsov and Paulos(2010)]%
        {Kuznetsov_2010_expertamateur}
\bibfield{author}{\bibinfo{person}{Stacey Kuznetsov} {and} \bibinfo{person}{Eric Paulos}.} \bibinfo{year}{2010}\natexlab{}.
\newblock \showarticletitle{Rise of the Expert Amateur: DIY Projects, Communities, and Cultures}. In \bibinfo{booktitle}{\emph{Proceedings of the 6th Nordic Conference on Human-Computer Interaction: Extending Boundaries}} \emph{(\bibinfo{series}{NordiCHI '10})}. \bibinfo{publisher}{ACM}, \bibinfo{pages}{295--304}.
\newblock
\showISBNx{978-1-60558-934-3}
\href{https://doi.org/10.1145/1868914.1868950}{doi:\nolinkurl{10.1145/1868914.1868950}}


\bibitem[Langlois et~al\mbox{.}(2016)]%
        {stochastic_structural_analysis}
\bibfield{author}{\bibinfo{person}{Timothy Langlois}, \bibinfo{person}{Ariel Shamir}, \bibinfo{person}{Daniel Dror}, \bibinfo{person}{Wojciech Matusik}, {and} \bibinfo{person}{David I.~W. Levin}.} \bibinfo{year}{2016}\natexlab{}.
\newblock \showarticletitle{Stochastic structural analysis for context-aware design and fabrication}.
\newblock \bibinfo{journal}{\emph{ACM Trans. Graph.}} \bibinfo{volume}{35}, \bibinfo{number}{6}, Article \bibinfo{articleno}{226} (\bibinfo{date}{dec} \bibinfo{year}{2016}), \bibinfo{numpages}{13}~pages.
\newblock
\showISSN{0730-0301}
\href{https://doi.org/10.1145/2980179.2982436}{doi:\nolinkurl{10.1145/2980179.2982436}}


\bibitem[Li et~al\mbox{.}(2023)]%
        {All-in-One-Print}
\bibfield{author}{\bibinfo{person}{Jiaji Li}, \bibinfo{person}{Mingming Li}, \bibinfo{person}{Junzhe Ji}, \bibinfo{person}{Deying Pan}, \bibinfo{person}{Yitao Fan}, \bibinfo{person}{Kuangqi Zhu}, \bibinfo{person}{Yue Yang}, \bibinfo{person}{Zihan Yan}, \bibinfo{person}{Lingyun Sun}, \bibinfo{person}{Ye Tao}, {et~al\mbox{.}}} \bibinfo{year}{2023}\natexlab{}.
\newblock \showarticletitle{All-in-one print: Designing and 3D printing dynamic objects using kinematic mechanism without assembly}. In \bibinfo{booktitle}{\emph{Proceedings of the 2023 CHI Conference on Human Factors in Computing Systems}}. \bibinfo{pages}{1--15}.
\newblock


\bibitem[Liao et~al\mbox{.}(2023)]%
        {liao2023interaction}
\bibfield{author}{\bibinfo{person}{Yi-Chi Liao}, \bibinfo{person}{John~J Dudley}, \bibinfo{person}{George~B Mo}, \bibinfo{person}{Chun-Lien Cheng}, \bibinfo{person}{Liwei Chan}, \bibinfo{person}{Antti Oulasvirta}, {and} \bibinfo{person}{Per~Ola Kristensson}.} \bibinfo{year}{2023}\natexlab{}.
\newblock \showarticletitle{Interaction Design With Multi-objective Bayesian Optimization}.
\newblock \bibinfo{journal}{\emph{IEEE Pervasive Computing}} \bibinfo{volume}{22}, \bibinfo{number}{1} (\bibinfo{year}{2023}), \bibinfo{pages}{29--38}.
\newblock


\bibitem[Louie et~al\mbox{.}(2020)]%
        {AI-music}
\bibfield{author}{\bibinfo{person}{Ryan Louie}, \bibinfo{person}{Andy Coenen}, \bibinfo{person}{Cheng~Zhi Huang}, \bibinfo{person}{Michael Terry}, {and} \bibinfo{person}{Carrie~J Cai}.} \bibinfo{year}{2020}\natexlab{}.
\newblock \showarticletitle{Novice-AI music co-creation via AI-steering tools for deep generative models}. In \bibinfo{booktitle}{\emph{Proceedings of the 2020 CHI conference on human factors in computing systems}}. \bibinfo{pages}{1--13}.
\newblock


\bibitem[Ma et~al\mbox{.}(2023)]%
        {x_mesh}
\bibfield{author}{\bibinfo{person}{Yiwei Ma}, \bibinfo{person}{Xiaoqing Zhang}, \bibinfo{person}{Xiaoshuai Sun}, \bibinfo{person}{Jiayi Ji}, \bibinfo{person}{Haowei Wang}, \bibinfo{person}{Guannan Jiang}, \bibinfo{person}{Weilin Zhuang}, {and} \bibinfo{person}{Rongrong Ji}.} \bibinfo{year}{2023}\natexlab{}.
\newblock \showarticletitle{X-Mesh: Towards Fast and Accurate Text-driven 3D Stylization via Dynamic Textual Guidance}. In \bibinfo{booktitle}{\emph{Proceedings of the IEEE/CVF International Conference on Computer Vision (ICCV)}}. \bibinfo{pages}{2749--2760}.
\newblock


\bibitem[Macklin(2022)]%
        {warp2022}
\bibfield{author}{\bibinfo{person}{Miles Macklin}.} \bibinfo{year}{2022}\natexlab{}.
\newblock \bibinfo{title}{Warp: A High-performance Python Framework for GPU Simulation and Graphics}.
\newblock \bibinfo{howpublished}{\url{https://github.com/nvidia/warp}}.
\newblock
\newblock
\shownote{NVIDIA GPU Technology Conference (GTC)}.


\bibitem[Mezghanni et~al\mbox{.}(2022)]%
        {mezghanni2022physical}
\bibfield{author}{\bibinfo{person}{Mariem Mezghanni}, \bibinfo{person}{Th{\'e}o Bodrito}, \bibinfo{person}{Malika Boulkenafed}, {and} \bibinfo{person}{Maks Ovsjanikov}.} \bibinfo{year}{2022}\natexlab{}.
\newblock \showarticletitle{Physical simulation layer for accurate 3d modeling}. In \bibinfo{booktitle}{\emph{Proceedings of the IEEE/CVF Conference on Computer Vision and Pattern Recognition}}. \bibinfo{pages}{13514--13523}.
\newblock


\bibitem[Mezghanni et~al\mbox{.}(2021)]%
        {mezghanni2021physically}
\bibfield{author}{\bibinfo{person}{Mariem Mezghanni}, \bibinfo{person}{Malika Boulkenafed}, \bibinfo{person}{Andre Lieutier}, {and} \bibinfo{person}{Maks Ovsjanikov}.} \bibinfo{year}{2021}\natexlab{}.
\newblock \showarticletitle{Physically-aware generative network for 3d shape modeling}. In \bibinfo{booktitle}{\emph{Proceedings of the IEEE/CVF Conference on Computer Vision and Pattern Recognition}}. \bibinfo{pages}{9330--9341}.
\newblock


\bibitem[Michel et~al\mbox{.}(2022)]%
        {michel2022text2mesh}
\bibfield{author}{\bibinfo{person}{Oscar Michel}, \bibinfo{person}{Roi Bar-On}, \bibinfo{person}{Richard Liu}, \bibinfo{person}{Sagie Benaim}, {and} \bibinfo{person}{Rana Hanocka}.} \bibinfo{year}{2022}\natexlab{}.
\newblock \showarticletitle{Text2mesh: Text-driven neural stylization for meshes}. In \bibinfo{booktitle}{\emph{Proceedings of the IEEE/CVF Conference on Computer Vision and Pattern Recognition}}. \bibinfo{pages}{13492--13502}.
\newblock


\bibitem[Mises(1913)]%
        {mises1913mechanik}
\bibfield{author}{\bibinfo{person}{R~v Mises}.} \bibinfo{year}{1913}\natexlab{}.
\newblock \showarticletitle{Mechanik der festen K{\"o}rper im plastisch-deformablen Zustand}.
\newblock \bibinfo{journal}{\emph{Nachrichten von der Gesellschaft der Wissenschaften zu G{\"o}ttingen, Mathematisch-Physikalische Klasse}}  \bibinfo{volume}{1913} (\bibinfo{year}{1913}), \bibinfo{pages}{582--592}.
\newblock


\bibitem[Norouzi et~al\mbox{.}(2021)]%
        {norouzi2021making}
\bibfield{author}{\bibinfo{person}{Behnaz Norouzi}, \bibinfo{person}{Marianne Kinnula}, {and} \bibinfo{person}{Netta Iivari}.} \bibinfo{year}{2021}\natexlab{}.
\newblock \showarticletitle{Making Sense of 3D Modelling and 3D Printing Activities of Young People: A Nexus Analytic Inquiry}. In \bibinfo{booktitle}{\emph{Proceedings of the 2021 CHI Conference on Human Factors in Computing Systems}} (Yokohama, Japan) \emph{(\bibinfo{series}{CHI '21})}. \bibinfo{publisher}{Association for Computing Machinery}, \bibinfo{address}{New York, NY, USA}, Article \bibinfo{articleno}{481}, \bibinfo{numpages}{16}~pages.
\newblock
\showISBNx{9781450380966}
\href{https://doi.org/10.1145/3411764.3445139}{doi:\nolinkurl{10.1145/3411764.3445139}}


\bibitem[Oehlberg et~al\mbox{.}(2015)]%
        {oehlberg2015patterns}
\bibfield{author}{\bibinfo{person}{Lora Oehlberg}, \bibinfo{person}{Wesley Willett}, {and} \bibinfo{person}{Wendy~E. Mackay}.} \bibinfo{year}{2015}\natexlab{}.
\newblock \showarticletitle{Patterns of Physical Design Remixing in Online Maker Communities}. In \bibinfo{booktitle}{\emph{Proceedings of the 33rd Annual ACM Conference on Human Factors in Computing Systems}} (Seoul, Republic of Korea) \emph{(\bibinfo{series}{CHI '15})}. \bibinfo{publisher}{Association for Computing Machinery}, \bibinfo{address}{New York, NY, USA}, \bibinfo{pages}{639–648}.
\newblock
\showISBNx{9781450331456}
\href{https://doi.org/10.1145/2702123.2702175}{doi:\nolinkurl{10.1145/2702123.2702175}}


\bibitem[Profita et~al\mbox{.}(2018)]%
        {Profita_hearing}
\bibfield{author}{\bibinfo{person}{Halley~P. Profita}, \bibinfo{person}{Abigale Stangl}, \bibinfo{person}{Laura Matuszewska}, \bibinfo{person}{Sigrunn Sky}, \bibinfo{person}{Raja Kushalnagar}, {and} \bibinfo{person}{Shaun~K. Kane}.} \bibinfo{year}{2018}\natexlab{}.
\newblock \showarticletitle{“Wear It Loud”: How and Why Hearing Aid and Cochlear Implant Users Customize Their Devices}.
\newblock \bibinfo{journal}{\emph{ACM Trans. Access. Comput.}} \bibinfo{volume}{11}, \bibinfo{number}{3}, Article \bibinfo{articleno}{13} (\bibinfo{date}{sep} \bibinfo{year}{2018}), \bibinfo{numpages}{32}~pages.
\newblock
\showISSN{1936-7228}
\href{https://doi.org/10.1145/3214382}{doi:\nolinkurl{10.1145/3214382}}


\bibitem[Radford et~al\mbox{.}(2021)]%
        {radford2021learning}
\bibfield{author}{\bibinfo{person}{Alec Radford}, \bibinfo{person}{Jong~Wook Kim}, \bibinfo{person}{Chris Hallacy}, \bibinfo{person}{Aditya Ramesh}, \bibinfo{person}{Gabriel Goh}, \bibinfo{person}{Sandhini Agarwal}, \bibinfo{person}{Girish Sastry}, \bibinfo{person}{Amanda Askell}, \bibinfo{person}{Pamela Mishkin}, \bibinfo{person}{Jack Clark}, \bibinfo{person}{Gretchen Krueger}, {and} \bibinfo{person}{Ilya Sutskever}.} \bibinfo{year}{2021}\natexlab{}.
\newblock \showarticletitle{Learning Transferable Visual Models From Natural Language Supervision}. In \bibinfo{booktitle}{\emph{Proceedings of the 38th International Conference on Machine Learning}} \emph{(\bibinfo{series}{Proceedings of Machine Learning Research}, Vol.~\bibinfo{volume}{139})}, \bibfield{editor}{\bibinfo{person}{Marina Meila} {and} \bibinfo{person}{Tong Zhang}} (Eds.). \bibinfo{publisher}{PMLR}, \bibinfo{pages}{8748--8763}.
\newblock
\urldef\tempurl%
\url{https://proceedings.mlr.press/v139/radford21a.html}
\showURL{%
\tempurl}


\bibitem[Rombach et~al\mbox{.}(2022)]%
        {rombach2022high}
\bibfield{author}{\bibinfo{person}{Robin Rombach}, \bibinfo{person}{Andreas Blattmann}, \bibinfo{person}{Dominik Lorenz}, \bibinfo{person}{Patrick Esser}, {and} \bibinfo{person}{Bj{\"o}rn Ommer}.} \bibinfo{year}{2022}\natexlab{}.
\newblock \showarticletitle{High-resolution image synthesis with latent diffusion models}. In \bibinfo{booktitle}{\emph{Proceedings of the IEEE/CVF conference on computer vision and pattern recognition}}. \bibinfo{pages}{10684--10695}.
\newblock


\bibitem[Roumen et~al\mbox{.}(2018)]%
        {roumen2018grafter}
\bibfield{author}{\bibinfo{person}{Thijs~Jan Roumen}, \bibinfo{person}{Willi M\"{u}ller}, {and} \bibinfo{person}{Patrick Baudisch}.} \bibinfo{year}{2018}\natexlab{}.
\newblock \showarticletitle{Grafter: Remixing 3D-Printed Machines}. In \bibinfo{booktitle}{\emph{Proceedings of the 2018 CHI Conference on Human Factors in Computing Systems}} (Montreal QC, Canada) \emph{(\bibinfo{series}{CHI '18})}. \bibinfo{publisher}{Association for Computing Machinery}, \bibinfo{address}{New York, NY, USA}, \bibinfo{pages}{1–12}.
\newblock
\showISBNx{9781450356206}
\href{https://doi.org/10.1145/3173574.3173637}{doi:\nolinkurl{10.1145/3173574.3173637}}


\bibitem[Schmidt and Ratto(2013)]%
        {schmidt2013design}
\bibfield{author}{\bibinfo{person}{Ryan Schmidt} {and} \bibinfo{person}{Matt Ratto}.} \bibinfo{year}{2013}\natexlab{}.
\newblock \showarticletitle{Design-to-Fabricate: Maker Hardware Requires Maker Software}.
\newblock \bibinfo{journal}{\emph{IEEE Computer Graphics and Applications}} \bibinfo{volume}{33}, \bibinfo{number}{6} (\bibinfo{year}{2013}), \bibinfo{pages}{26--34}.
\newblock
\href{https://doi.org/10.1109/MCG.2013.90}{doi:\nolinkurl{10.1109/MCG.2013.90}}


\bibitem[Shinohara et~al\mbox{.}(2018)]%
        {Shinohara_social}
\bibfield{author}{\bibinfo{person}{Kristen Shinohara}, \bibinfo{person}{Cynthia~L. Bennett}, \bibinfo{person}{Wanda Pratt}, {and} \bibinfo{person}{Jacob~O. Wobbrock}.} \bibinfo{year}{2018}\natexlab{}.
\newblock \showarticletitle{Tenets for Social Accessibility: Towards Humanizing Disabled People in Design}.
\newblock \bibinfo{journal}{\emph{ACM Trans. Access. Comput.}} \bibinfo{volume}{11}, \bibinfo{number}{1}, Article \bibinfo{articleno}{6} (\bibinfo{date}{mar} \bibinfo{year}{2018}), \bibinfo{numpages}{31}~pages.
\newblock
\showISSN{1936-7228}
\href{https://doi.org/10.1145/3178855}{doi:\nolinkurl{10.1145/3178855}}


\bibitem[Shugrina et~al\mbox{.}(2015)]%
        {shugrina2015fab}
\bibfield{author}{\bibinfo{person}{Maria Shugrina}, \bibinfo{person}{Ariel Shamir}, {and} \bibinfo{person}{Wojciech Matusik}.} \bibinfo{year}{2015}\natexlab{}.
\newblock \showarticletitle{Fab Forms: Customizable Objects for Fabrication with Validity and Geometry Caching}.
\newblock \bibinfo{journal}{\emph{ACM Trans. Graph.}} \bibinfo{volume}{34}, \bibinfo{number}{4}, Article \bibinfo{articleno}{100} (\bibinfo{date}{jul} \bibinfo{year}{2015}), \bibinfo{numpages}{12}~pages.
\newblock
\showISSN{0730-0301}
\href{https://doi.org/10.1145/2766994}{doi:\nolinkurl{10.1145/2766994}}


\bibitem[Skouras et~al\mbox{.}(2013)]%
        {skouras2013computational}
\bibfield{author}{\bibinfo{person}{M{\'e}lina Skouras}, \bibinfo{person}{Bernhard Thomaszewski}, \bibinfo{person}{Stelian Coros}, \bibinfo{person}{Bernd Bickel}, {and} \bibinfo{person}{Markus Gross}.} \bibinfo{year}{2013}\natexlab{}.
\newblock \showarticletitle{Computational design of actuated deformable characters}.
\newblock \bibinfo{journal}{\emph{ACM Transactions on Graphics (TOG)}} \bibinfo{volume}{32}, \bibinfo{number}{4} (\bibinfo{year}{2013}), \bibinfo{pages}{1--10}.
\newblock


\bibitem[Smith et~al\mbox{.}(2018)]%
        {smith2018stable}
\bibfield{author}{\bibinfo{person}{Breannan Smith}, \bibinfo{person}{Fernando~De Goes}, {and} \bibinfo{person}{Theodore Kim}.} \bibinfo{year}{2018}\natexlab{}.
\newblock \showarticletitle{Stable neo-hookean flesh simulation}.
\newblock \bibinfo{journal}{\emph{ACM Transactions on Graphics (TOG)}} \bibinfo{volume}{37}, \bibinfo{number}{2} (\bibinfo{year}{2018}), \bibinfo{pages}{1--15}.
\newblock


\bibitem[Stava et~al\mbox{.}(2012)]%
        {stress_relief}
\bibfield{author}{\bibinfo{person}{Ondrej Stava}, \bibinfo{person}{Juraj Vanek}, \bibinfo{person}{Bedrich Benes}, \bibinfo{person}{Nathan Carr}, {and} \bibinfo{person}{Radom\'{\i}r M\v{e}ch}.} \bibinfo{year}{2012}\natexlab{}.
\newblock \showarticletitle{Stress relief: improving structural strength of 3D printable objects}.
\newblock \bibinfo{journal}{\emph{ACM Trans. Graph.}} \bibinfo{volume}{31}, \bibinfo{number}{4}, Article \bibinfo{articleno}{48} (\bibinfo{date}{jul} \bibinfo{year}{2012}), \bibinfo{numpages}{11}~pages.
\newblock
\showISSN{0730-0301}
\href{https://doi.org/10.1145/2185520.2185544}{doi:\nolinkurl{10.1145/2185520.2185544}}


\bibitem[Sun et~al\mbox{.}(2022)]%
        {X-Bridges}
\bibfield{author}{\bibinfo{person}{Lingyun Sun}, \bibinfo{person}{Jiaji Li}, \bibinfo{person}{Junzhe Ji}, \bibinfo{person}{Deying Pan}, \bibinfo{person}{Mingming Li}, \bibinfo{person}{Kuangqi Zhu}, \bibinfo{person}{Yitao Fan}, \bibinfo{person}{Yue Yang}, \bibinfo{person}{Ye Tao}, {and} \bibinfo{person}{Guanyun Wang}.} \bibinfo{year}{2022}\natexlab{}.
\newblock \showarticletitle{X-Bridges: Designing Tunable Bridges to Enrich 3D Printed Objects' Deformation and Stiffness}. In \bibinfo{booktitle}{\emph{Proceedings of the 35th Annual ACM Symposium on User Interface Software and Technology}} (Bend, OR, USA) \emph{(\bibinfo{series}{UIST '22})}. \bibinfo{publisher}{Association for Computing Machinery}, \bibinfo{address}{New York, NY, USA}, Article \bibinfo{articleno}{20}, \bibinfo{numpages}{12}~pages.
\newblock
\showISBNx{9781450393201}
\href{https://doi.org/10.1145/3526113.3545710}{doi:\nolinkurl{10.1145/3526113.3545710}}


\bibitem[Wang et~al\mbox{.}(2024)]%
        {wang2024diffusebot}
\bibfield{author}{\bibinfo{person}{Tsun-Hsuan~Johnson Wang}, \bibinfo{person}{Juntian Zheng}, \bibinfo{person}{Pingchuan Ma}, \bibinfo{person}{Yilun Du}, \bibinfo{person}{Byungchul Kim}, \bibinfo{person}{Andrew Spielberg}, \bibinfo{person}{Josh Tenenbaum}, \bibinfo{person}{Chuang Gan}, {and} \bibinfo{person}{Daniela Rus}.} \bibinfo{year}{2024}\natexlab{}.
\newblock \showarticletitle{DiffuseBot: Breeding Soft Robots With Physics-Augmented Generative Diffusion Models}.
\newblock \bibinfo{journal}{\emph{Advances in Neural Information Processing Systems}}  \bibinfo{volume}{36} (\bibinfo{year}{2024}).
\newblock


\bibitem[Yan et~al\mbox{.}(2021)]%
        {yan2021man}
\bibfield{author}{\bibinfo{person}{Xin Yan}, \bibinfo{person}{Lin Lu}, \bibinfo{person}{Andrei Sharf}, \bibinfo{person}{Xing Yu}, {and} \bibinfo{person}{Yulu Sun}.} \bibinfo{year}{2021}\natexlab{}.
\newblock \showarticletitle{Man-made by computer: On-the-fly fine texture 3D printing}. In \bibinfo{booktitle}{\emph{Proceedings of the 6th Annual ACM Symposium on Computational Fabrication}}. \bibinfo{pages}{1--10}.
\newblock


\bibitem[Zehnder et~al\mbox{.}(2016)]%
        {zehnder2016designing}
\bibfield{author}{\bibinfo{person}{Jonas Zehnder}, \bibinfo{person}{Stelian Coros}, {and} \bibinfo{person}{Bernhard Thomaszewski}.} \bibinfo{year}{2016}\natexlab{}.
\newblock \showarticletitle{Designing structurally-sound ornamental curve networks}.
\newblock \bibinfo{journal}{\emph{ACM Transactions on Graphics (TOG)}} \bibinfo{volume}{35}, \bibinfo{number}{4} (\bibinfo{year}{2016}), \bibinfo{pages}{1--10}.
\newblock


\bibitem[Zhou et~al\mbox{.}(2013)]%
        {zhou_worst_case_structural}
\bibfield{author}{\bibinfo{person}{Qingnan Zhou}, \bibinfo{person}{Julian Panetta}, {and} \bibinfo{person}{Denis Zorin}.} \bibinfo{year}{2013}\natexlab{}.
\newblock \showarticletitle{Worst-case structural analysis}.
\newblock \bibinfo{journal}{\emph{ACM Trans. Graph.}} \bibinfo{volume}{32}, \bibinfo{number}{4}, Article \bibinfo{articleno}{137} (\bibinfo{date}{jul} \bibinfo{year}{2013}), \bibinfo{numpages}{12}~pages.
\newblock
\showISSN{0730-0301}
\href{https://doi.org/10.1145/2461912.2461967}{doi:\nolinkurl{10.1145/2461912.2461967}}


\end{thebibliography}
